\documentclass[useAMS,usenatbib]{mn2e}

\usepackage{amssymb,natbib,graphicx,color,ctable}
\usepackage{longtable}
\usepackage{threeparttablex}
\usepackage{wasysym}
\usepackage{url}

\title[Kinematics and dark matter fractions of $z=2$ galaxies in TNG50]{The Kinematics and Dark Matter Fractions of TNG50 Galaxies at $z=2$ from an Observational Perspective}

\author[{\"U}bler, H. et al.]
{\parbox{20cm}{
Hannah {\"U}bler$^{1}$\thanks{E-mail: hannah@mpe.mpg.de},
Shy Genel$^{2,3}$,
Amiel Sternberg$^{4,2,1}$,
Reinhard Genzel$^{1,5}$,\\
Sedona H.~Price$^{1}$,
Natascha M.\ F\"orster Schreiber$^{1}$,
Taro T.~Shimizu$^{1}$,\\
Annalisa Pillepich$^{6}$,
Dylan Nelson$^{7}$,
Andreas Burkert$^{1,8}$,
Ric Davies$^{1}$,\\
Lars Hernquist$^{9}$,
Philipp Lang$^{6}$,
Dieter Lutz$^{1}$,
R{\"u}diger Pakmor$^{7}$,\\
and Linda J.~Tacconi$^{1}$}\vspace{0.3cm}\\
$^{1}$Max-Planck-Institut f{\"u}r extraterrestrische Physik, Gie\ss enbachstra\ss e 1, 85748 Garching bei M{\"u}nchen, Germany\\
$^{2}$Center for Computational Astrophysics, Flatiron Institute, 162 Fifth Avenue, New York, NY 10010, USA\\
$^{3}$Columbia Astrophysics Laboratory, Columbia University, 550 West 120th Street, New York, NY 10027, USA\\
$^{4}$School of Physics \& Astronomy, Tel Aviv University, Ramat Aviv 69978, Israel\\
$^{5}$Departments of Physics and Astronomy, University of California, Berkeley, CA 94720, USA\\
$^{6}$Max-Planck-Institut f{\"u}r Astronomie, K{\"o}nigstuhl 17, 69117 Heidelberg, Germany\\
$^{7}$Max-Planck-Institut f\"ur Astrophysik, Karl Schwarzschildstr.\ 1, D-85737 Garching, Germany\\
$^{8}$Universit\"ats-Sternwarte, Ludwig-Maximilians-Universit\"at M\"unchen, Scheinerstr. 1, 81679 M\"unchen, Germany\\
$^{9}$Harvard-Smithsonian Center for Astrophysics, 60 Garden Street, Cambridge, MA 02138, USA\\
}

\begin{document}

\maketitle

\label{firstpage}

\begin{abstract}
We contrast the gas kinematics and dark matter contents of $z=2$ star-forming galaxies (SFGs) from state-of-the-art cosmological simulations within the $\Lambda$CDM framework to observations. To this end, we create realistic mock observations of massive SFGs ($M_*>4\times10^{10} M_{\odot}$, SFR~$>50~M_{\odot}$ yr$^{-1}$) from the TNG50 simulation of the IllustrisTNG suite, resembling near-infrared, adaptive-optics assisted integral-field observations from the ground.
Using observational line fitting and modeling techniques, we analyse in detail the kinematics of seven TNG50 galaxies from five different projections per galaxy, and compare them to observations of twelve massive SFGs by \cite{Genzel20}.
The simulated galaxies show clear signs of disc rotation but mostly exhibit more asymmetric rotation curves, partly due to large intrinsic radial and vertical velocity components. At identical inclination angle, their one-dimensional velocity profiles can vary along different lines of sight by up to $\Delta v=200$~km~s$^{-1}$. 
From dynamical modelling we infer rotation speeds and velocity dispersions that are broadly consistent with observational results. 
We find low central dark matter fractions compatible with observations ($f_{\rm DM}^v(<R_e)=v_{\rm DM}^2(R_e)/v_{\rm circ}^2(R_e)\sim0.32\pm0.10$), however for disc effective radii $R_e$ that are mostly too small: at fixed $R_e$ the TNG50 dark matter fractions are too high by a factor of $\sim2$. 
We speculate that the differences in gas kinematics and dark matter content compared to the observations may be due to physical processes that are not resolved in sufficient detail with the numerical resolution available in current cosmological simulations.
\end{abstract}

\begin{keywords}
methods: numerical -- galaxies: high-redshift -- galaxies: kinematics and dynamics
\end{keywords}

\section{Introduction}\label{s:intro}

Recent observations of massive ($M_*\approx 10^{11} M_\odot$) star-forming galaxies (SFGs) at redshift $z\sim2$, near the peak of cosmic star-formation rate density, have demonstrated that these rapidly evolving galaxies differ from present-day systems in several fundamental ways. First, the $z\sim2$ SFGs have higher gas-to-stellar mass ratios ($M_{\rm gas}/M_*\sim1$; \citealp[e.g.][]{Genzel15, Scoville17, Tacconi18}). Second, they are forming stars more rapidly (SFR/$M_*\sim1.2$ Gyr$^{-1}$; \citealp[e.g.][]{Daddi07, Rodighiero11, Whitaker14, Speagle14}). Third, they have higher intrinsic velocity dispersions relative to ordered rotational motions ($\sigma_0/v_{\rm rot}\sim 0.2$; \citealp[e.g.][]{FS06b, Genzel08, Genzel11, Wisnioski15, Simons17}).

In addition to these differences in global properties, several kinematic studies of individual galaxies have also revealed that the central regions of these most massive $z\sim2$ SFGs are strongly baryon-dominated \citep[e.g.][]{Alcorn16, Price16, Price19, WuytsS16, Genzel17, Genzel20}, with galaxy-scale dark matter fractions much lower than for typical SFGs at the current epoch \citep[e.g.][]{Martinsson13a, Martinsson13b}. 
Because studies of resolved properties are observationally very expensive, stacking approaches have been used to determine typical dynamical properties of high$-z$ SFGs. However, current stacking studies disagree on their main results, largely driven by different methodological concepts and possibly selection effects \citep{Lang17, Tiley19}. 
The differences should not be over-interpreted: the results of dynamical studies of individual galaxies in the local Universe \citep[e.g.][]{Persic88, Begeman91, Sancisi04, Noordermeer07b, deBlok08, Lelli16} as well as at high redshift \citep{Genzel17, Genzel20, Uebler18, Lelli18, Motta18, Drew18} show that both rotation curve shapes and dynamical support are contingent on other galaxy properties, such as velocity dispersion, baryonic mass, baryonic surface density, or bulge mass, properties not systematically controlled for in current stacking analyses. Therefore, kinematic studies of individual galaxies still constitute the most robust reference.

Reproducing the detailed properties of high$-z$ galaxies poses a challenge to simulations \citep[see review by][]{Naab17}. 
To make progress, recent studies are now focusing on specific tests of simulations against kinematic data by means of mock observations, with varying degrees of observational realism \citep[e.g.][]{Genel12a, Lovell18, Teklu18, Pillepich19, Simons19, Wellons19}. 
As one of the most recent models, the IllustrisTNG simulation suite \citep{NelsonD18, Naiman18, Marinacci18, Pillepich18, Springel18} provides several large realizations of cosmological galaxy populations that can be compared with data. 

The goal of this paper is to contrast simulated SFGs from the highest-resolution run of the IllustrisTNG suite, TNG50 \citep{NelsonD19, Pillepich19}, to a subsample of recent detailed observations by \cite{Genzel20}. Our main focus is on the kinematics and the associated dark-matter distributions of the most massive $z\sim2$ SFGs. For this purpose we create realistic mock observations of the star-forming gas kinematics of selected massive $z=2$ galaxies from TNG50, including specific instrumental effects, and random as well as systematic noise affecting near-infrared observations from the ground. We then apply the same data extraction pipeline and modeling tools to the simulated galaxies that were applied to the galaxies by \cite{Genzel20}. 

This approach enables two types of comparisons. 
First, since the internal structures of the simulated galaxies can be inspected directly, we can assess how accurately the observational pipelines recover (complex) intrinsic structures. 
Second, given the observational results, particularly the low central dark matter fractions and the role of pressure support as frequently indicated by declining rotation curves, we can ask whether the IllustrisTNG model successfully reproduces the observed properties, or not, assuming that we can identify simulated analogs from the TNG50 volume that are similar enough to the observed galaxies.  

This paper is structured as follows: in Section~\ref{s:obs}, we briefly describe the selection, modeling assumptions, and interpretation of our observational comparison sample by \cite{Genzel20}. In Section~\ref{s:methods}, we discuss the selection of galaxies from the TNG50 simulation, our mock observations, kinematic analysis, and modeling. In Section~\ref{s:results}, we present both mock-observed and intrinsic kinematics, modeling results, and compare to the observational sample with a focus on galaxy-scale dark matter fractions. We summarize and conclude in Section~\ref{s:conclusions}.

\section{The observational picture}\label{s:obs}

The pioneering work by \cite{Genzel17} revealed declining rotation curves for a sample of six massive, extended SFGs at $0.9<z<2.4$. Through dynamical modeling of these deep and high-quality data, it was possible to estimate the central dark matter fractions based on a standard Navarro, Frenk, \& White halo model \citep[NFW;][]{NFW96}. This analysis showed that the galaxies had low to negligible central dark matter fractions within the baryonic disc effective radius $R_e$.\footnote{
Throughout the paper, we use $R_e$ to refer to the baryonic half-mass radius of the thick exponential disc component constrained through dynamical modeling. We use $R_{1/2}$ to refer to the total half-light radius (including the bulge).}
This first study was substantially enlarged through the recent work by \cite{Genzel20}, presenting modeling and analysis of 41 galaxies (including the objects presented by \citealp{Genzel17}), with a focus on extending towards lower masses and redshifts. This diversified view on the high$-z$ SFG population reveals a variety of kinematic and dark matter properties.

\subsection{Observational comparison sample}

The \cite{Genzel20} galaxies were selected from the SINS/zC-SINF and KMOS$^{\rm 3D}$ integral-field spectroscopic surveys \citep{FS09, FS18, Wisnioski15, Wisnioski19}, and from the PHIBSS 1 \& 2 interferometric surveys \citep{Tacconi10, Tacconi13, Tacconi18, Freundlich19}. This selection was based on the quality of the available data, extended galaxy sizes ($R_{1/2}\gtrsim2$~kpc), and sufficiently high H$\alpha$ or CO surface brightness \citep[see][for more details]{Genzel17, Genzel20}.

Among these 41 galaxies, the population of the most massive SFGs at redshift $z\sim2$ is especially interesting because of their low dark matter fractions measured at their dynamically inferred disc effective radii, despite having fairly extended discs. Reproducing galaxies with these physical properties appears particularly challenging for current theoretical models of galaxy formation, including the IllustrisTNG model \citep{Lovell18}.\footnote{
Low central dark matter fractions have been intrinsically measured for more compact simulated high$-z$ SFGs at smaller radii, e.g.\ $\leq30$ per cent for SFGs at stellar half-mass radii of $\sim2$~kpc in the Magneticum Pathfinder simulations \citep{Teklu18}, or $\sim8$ per cent at stellar half-mass radii of $\sim1$~kpc in the FIRE-2 model \citep{Wellons19}.}
Therefore, we focus in the present work on comparing TNG50 galaxies to the subsample of observed galaxies at $z\geq1.5$ with stellar masses $M_*>4\times 10^{10}M_{\odot}$. We choose this mass cut to include the five $z\geq1.5$ galaxies first presented by \cite{Genzel17}. Employing the same cuts on the larger sample recently published by \cite{Genzel20} supplies another seven galaxies, for a total comparison sample of twelve observed galaxies. Throughout this work, we refer to these twelve galaxies as the selected G20 sample.

The data for galaxies in our selected G20 sample are adaptive-optics assisted and/or seeing-limited, and one galaxy additionally has interferometric observations. For our main mock analysis of the TNG50 galaxies (see Section~\ref{s:mock}), we focus on adaptive-optics assisted data quality representing the majority of observations by \cite{Genzel17}.

As shown in the left panel of Figure~\ref{fig:selection}, our selected G20 galaxies lie along or somewhat above the main sequence of SFGs at their respective redshifts, with star formation rates SFR=$50-400~M_{\odot}$~yr$^{-1}$. Similarly, their half-light radii ($R_{1/2}\sim3-9$~kpc) lie along or somewhat above the mass-size relation (middle panel). Their stellar masses are in the range $4\times10^{10}<M_*/M_{\odot}<3.2\times10^{11}$.
All galaxies have circular velocities $v_{\rm circ}(R_e)\sim250-420$~km~s$^{-1}$ and intrinsic velocity dispersions $\sigma_0\sim20-80$~km~s$^{-1}$.
The dynamical analysis by \cite{Genzel20} shows that these massive, high$-z$ SFGs are baryon-dominated within $R_e$, with $f_{\rm DM}^v(<R_e)=v_{\rm DM}^2(R_e)/v_{\rm circ}^2(R_e)\lesssim0.4$, where $v_{\rm circ}$ is the total circular velocity, and $v_{\rm DM}$ is the velocity due to dark matter.

\subsection{Dynamical modeling assumptions}

For modeling the baryonic mass distribution, \cite{Genzel20}, following \cite{Genzel17}, considered a combination of a thick exponential and axisymmetric disc and a compact bulge. These choices were motivated by the typical structural properties of high$-z$ SFGs \citep{WuytsS11a, vdWel14b, Lang14}, and the available ancillary data.
For the results on the dark matter distribution we quote in this paper, \cite{Genzel20} adopted an NFW halo profile.
The NFW profile is a two-power-law density model of the form
\begin{equation}\label{eq:twopower}
	\rho(r) = \frac{\rho_0}{(r/r_s)^\alpha (1+r/r_s)^{\beta-\alpha}},
\end{equation}
where $\alpha=1$ and $\beta=3$. In this expression, $r_s$ is the halo scale radius, and $\rho_0$ is the characteristic dark matter density. 
The halo concentration parameter $c\equiv R_{200,c}/r_s$, where $R_{200,c}$ is the virial radius within which the mean enclosed density equals 200 times the critical density of the Universe. The halo concentration parameter is set to typical values determined through the estimated halo mass based on the stellar mass and redshift of each galaxy \citep{Moster13} and assuming standard concentration-mass relations from dark matter-only simulations \citep{Dutton14}.
\cite{Genzel20} performed fits with and without adiabatic contraction being effective \citep[see][]{Blumenthal86}. Their quoted modeling results which we use in this paper, specifically for the central dark matter fraction, represent averages of fits with and without adiabatic contraction (see their Tables D1 and D2).

The dynamical analysis further accounted for pressure support expected from turbulent motions following \cite{Burkert10, Burkert16}. This correction results in a reduction of the rotation velocity $v_{\rm rot}$ with respect to the circular velocity $v_{\rm circ}$ as a function of radius:
\begin{equation}\label{eq:pressure}
    v_{\rm rot}(r)\equiv \sqrt{v_{\rm circ}^2(r)-2\sigma_0^2\frac{r}{R_d}},
\end{equation}
where $R_d$ is the disc scale length, and the velocity dispersion $\sigma_0$ is assumed to be constant throughout the disc.

We adopt these basic assumptions for our modeling of the simulated galaxies, as detailed in Section~\ref{s:modeling}. However, differently than \cite{Genzel20}, we do not average NFW fits with and without adiabatic contraction, but instead consider pure NFW haloes together with contracted dark matter haloes based on their intrinsic mass distributions.

\subsection{Observational interpretation}\label{s:obsinterp}

The prominent drop in the observed rotation curves of the majority of the most massive, high$-z$ SFGs (see Figures 1 and 2 by \citealp{Genzel17}, and Figure 4 by \citealp{Genzel20}) can be explained by a combination of two effects: (i) low central dark matter fractions, and (ii) high turbulent motions that produce outward pressure gradients that counteract inward gravity, leading to reduced rotational speeds at large radii. 

The central dark matter fractions inferred by \cite{Genzel20} for the massive $z\geq1.5$ SFGs we consider in this work are typically lower than predicted from abundance matching in conjunction with NFW halo profiles \citep{Moster13, Moster18, Behroozi13b}, and also in comparison to lower resolution cosmological simulations such as TNG100 (\citealp{Lovell18}; see also Figure 6 by \citealp{Genzel20}). 
Potential reasons for this are small-scale physical processes that might not be adequately captured in large-scale simulations, particularly at early cosmic times: (i) high$-z$ SFGs are more gas-rich than their equal-mass $z=0$ counterparts, with dissipation processes efficiently channeling baryonic material to the central regions. (ii) Dark matter could be removed from the central galactic regions due to strong AGN and/or stellar feedback, for which there is clear evidence from observations of gas outflows at high redshift \citep[e.g.][and reference therein]{Shapley03, Weiner09, Genzel11, Genzel14, Harrison16, FS19, Freeman19, Swinbank19}, or due to heating of the halo via dynamical friction caused by in-spiraling baryonic material (e.g.\ \citealp{ElZant01, Martizzi13, Freundlich20a}; A.~Burkert et al., in prep.). A consequence could be an alteration of the dark matter density profiles with less dense cores, such as \cite{Burkert95} profiles or certain \cite{Einasto65} profile solutions.

Based on a comparison of the mass budget in local galaxies with the high baryonic masses already assembled in the high$-z$ SFGs, \cite{Genzel17, Genzel20} concluded that their results are consistent with high$-z$ SFGs likely evolving into early-type systems by the present day, after further consumption and/or ejection of their available cold gas. Since present-day early-type galaxies have similarly low central dark matter fractions \citep[e.g.][]{Thomas11, Cappellari13, Mendel20}, this suggests that the central mass budget is set early on in the evolution of the most massive galaxies.

\begin{figure*}
    \centering
    \includegraphics[width=0.33\textwidth]{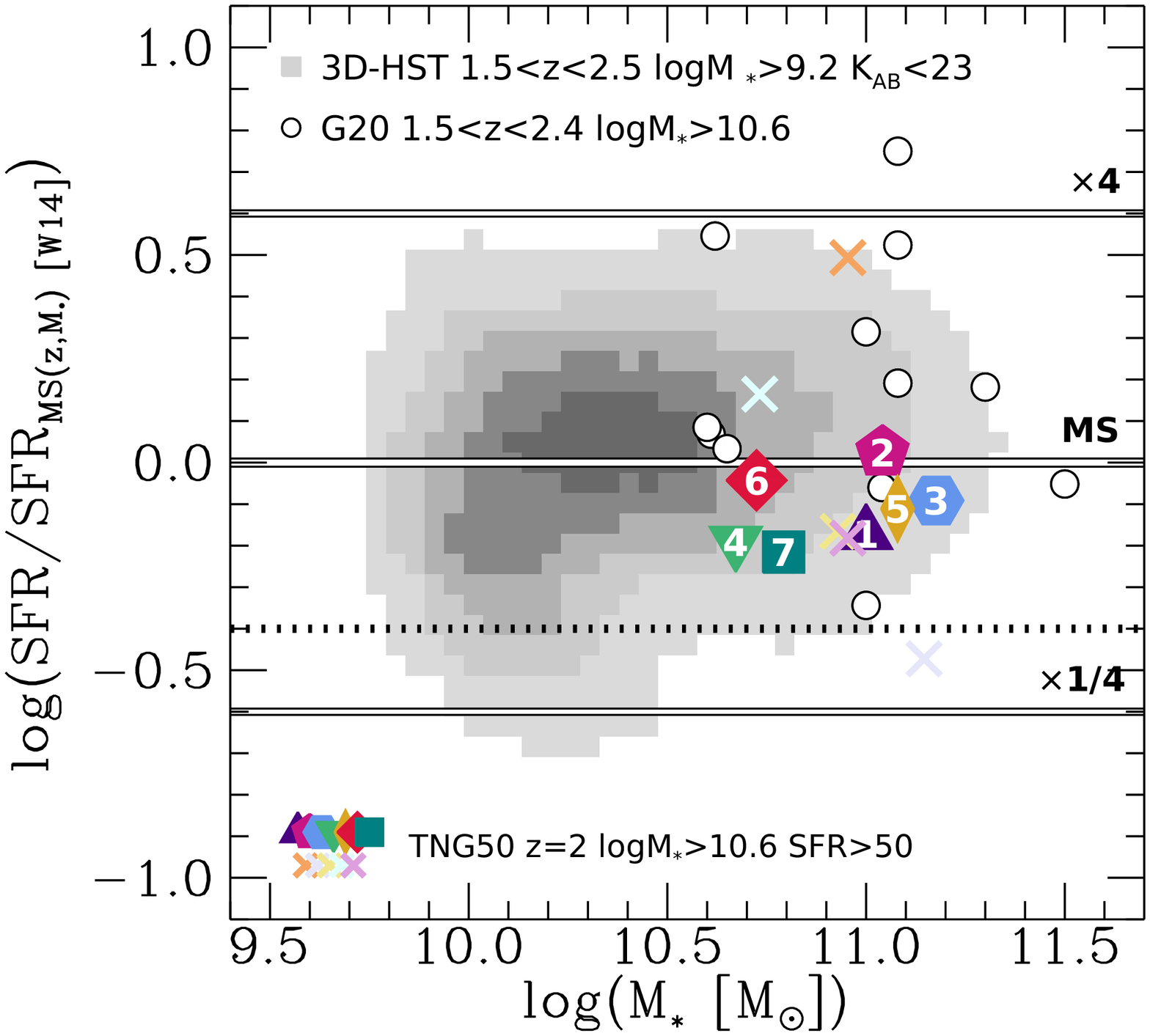}
    \includegraphics[width=0.33\textwidth]{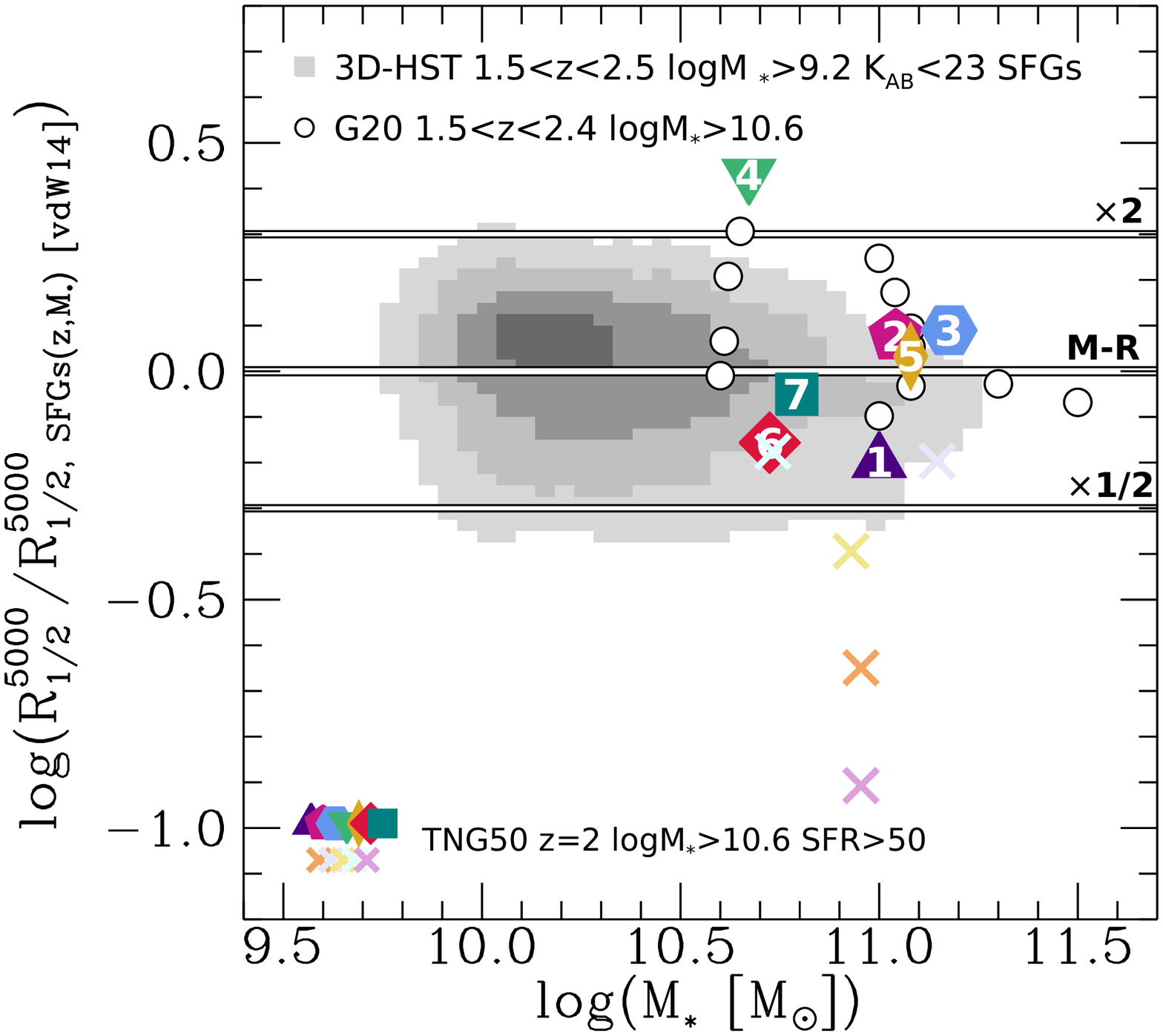}
    \includegraphics[width=0.33\textwidth]{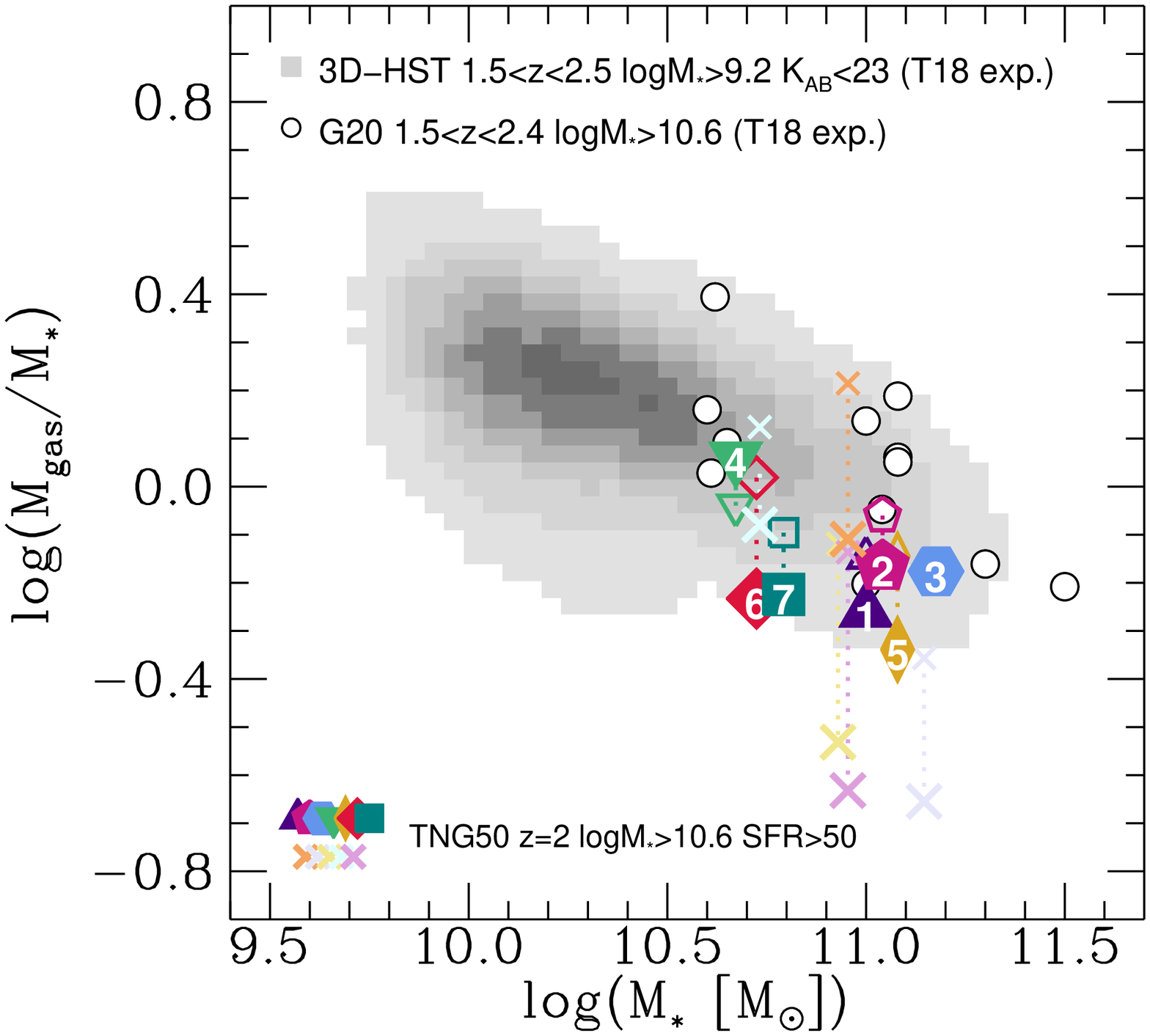}
    \caption{Intrinsic physical properties of TNG50 galaxies selected from the $z=2$ snapshot with $M_* > 4 \times 10^{10} M_{\odot}$ and SFR $> 50 M_{\odot}$~yr$^{-1}$ (colored symbols; cf.\ Table~\ref{tab:tngprop}) in comparison to the observational reference sample by \citet{Genzel20} (G20; white circles), and to the underlying observed galaxy population at $1.5<z<2.5$ based on the 3D-HST catalogue (grey scale; \citealp{Brammer12, Skelton14, Momcheva16}). For the 3D-HST sample we apply the following cuts: log$(M_*/M_{\odot})>9.2$, $K_{\rm AB}<23$~mag, and for the middle panel also SFR$/M_*>0.7/t_{\rm Hubble}$. TNG50 galaxies that satisfy the stellar mass and SFR cuts but that are not included in the kinematic analysis (because undergoing a merger, disturbed or too compact, see main text) are shown as colored crosses. Masses and SFRs of the TNG50 galaxies are measured within a three-dimensional aperture with radius $20$~kpc around the potential minimum. Offset from the main sequence (MS; left), offset from the mass--size relation ($M-R$; middle), and gas-to-stellar mass ratio (right) are shown as a function of stellar mass. 
    In the left panel, the SFR is normalized to the main sequence as derived by \citet{Whitaker14} at the redshift and stellar mass of each galaxy, using the redshift-interpolated parametrization by \citet{Wisnioski15}. The dotted line indicates the approximate offset between the observationally calibrated main sequence, and the main sequence normalization based on TNG galaxies alone \citep{Donnari19err}. In the middle panel, the half-light sizes are corrected to the rest-frame 5000~\AA\, and normalized to the $M-R$ relation of SFGs as derived by \citet{vdWel14a} at the redshift and stellar mass of each galaxy. In the right panel, gas masses for the 3D-HST galaxies and the G20 sample are estimated from the scaling relations by \citeauthor{Tacconi18} (2018; T18). The smaller (open colored) symbols connected to the larger (filled colored) symbols indicate the values that would be expected for the TNG50 galaxies based on the T18 scaling relation.}
    \label{fig:selection}
\end{figure*}

\section{Simulated galaxies and methodology}
\label{s:methods}

\subsection{The TNG50 simulation}
\label{s:simulations}

The TNG50 simulation \citep{NelsonD19, Pillepich19} is the highest-resolution volume of the IllustrisTNG project \citep{NelsonD18, Naiman18, Marinacci18, Pillepich18, Springel18}, with a uniform periodic-boundary cube of 51.7 co-moving Mpc on a side, and $2\times2160^3$ initial resolution elements, half dark matter particles and half gas cells. The simulations are run with the unstructured moving-mesh code {\sc Arepo} \citep{Springel10} and incorporate dark matter, gas, stars, black holes, and magnetic fields. The dark matter and baryonic mass resolutions in TNG50 are $4.5\times10^5M_{\odot}$ and $8.5\times10^4M_{\odot}$, respectively, and the gravitational softening lengths at $z=2$ are 192~pc for stars and dark matter, and adaptive for gas, with a typical size of $100-200$~pc for star-forming gas. 
The simulations account for star formation, stellar population evolution, chemical enrichment through supernovae type Ia and II and through AGB stars, gas radiative processes, the formation, coalescence, and growth of supermassive black holes, and feedback from supernovae and black holes \citep{Weinberger17, Pillepich18a}. TNG50 adopts a \cite{Planck15xiii} cosmology with $h=0.68$, $\Omega_b=0.05$, $\Omega_m=0.31$, $\Omega_\Lambda=0.69$, and $\sigma_8=0.82$.

\subsection{Sample selection and global properties}
\label{s:selection}

To select simulated galaxies that resemble the available deep observational data of the most massive, high$-z$ SFGs, we choose central galaxies with stellar mass and SFR matched to the observed sample, with $M_* > 4\times10^{10} M_{\odot}$ and SFR~$>50M_{\odot}$~yr$^{-1}$ at $z=2$.
For this initial selection we consider the instantaneous SFR and stellar mass within twice the radius enclosing half of the gravitationally bound stellar mass.

It is known that the TNG model predicts SFRs that are systematically lower by $\sim0.4$~dex at $z\sim2$ compared to the \cite{Whitaker14} observational reference main sequence \citep{Donnari19, Donnari19err}, as indicated by the dotted line in the left panel of Figure~\ref{fig:selection}. Similar offsets relative to observations appear in many cosmological hydrodynamical simulations \citep[e.g.][]{Furlong15}, and are not yet fully understood \citep[but see the tensions between the observed evolution of galaxy masses or luminosity functions and the observed specific SFRs pointed out by e.g.][]{Leja15, Yu16, Wilkins19}. 
From the observational side, recent work indicates that masses (SFRs) based on SED-modeling are systematically underestimated (overestimated), however for high stellar masses and SFRs and at $z\sim2$, as in our sample, these effects are supposedly minor \citep{Leja19}. Therefore, in this paper, we proceed by selecting analogs of observed galaxies from the TNG50 galaxy population by imposing cuts on stellar mass and star formation rates based on limits taken at face value and without accounting for observational uncertainties. 

In total, 12 central galaxies in the TNG50 volume meet these cuts at $z=2$.
Of those, we further exclude five galaxies that are either very compact, therefore hampering the extraction of (resolved) kinematics out to sufficiently large radii, or that are clearly interacting or disturbed. We show projected kinematic maps of the dismissed galaxies in Appendix~\ref{s:dismissed}. 
The remaining seven galaxies are used for the kinematic analysis in this paper.

\begin{table}
\caption[Physical properties of IllustrisTNG50 galaxies]{Physical properties of the TNG50 galaxies selected for kinematic analysis (top seven rows), and excluded (bottom 5 rows): stellar mass $M_*$, gas mass $M_{\rm gas}$, and instantaneous star formation rate SFR, all within a three-dimensional aperture with radius 20~kpc around the potential minimum, and the three-dimensional rest-frame $V-$band half-light size $R_{1/2}$.}
\label{tab:tngprop}
\centering
\begin{tabular}{lcccc}
\hline
	ID, subhalo & $M_*$ & $M_{\rm gas}$ & SFR & $R_{1/2}$ \\ 
	  (symbol) & [$10^{11} M_{\odot}$] & [$10^{11} M_{\odot}$] & [$M_{\odot}~yr^{-1}$] & [kpc] \\ 
\hline
    \#1, 25822	($\bigtriangleup$) & $1.0$	& $0.6$	& 71    & 2.6 \\
	\#2, 39746	($\pentagon$) & $1.1$	& $0.8$ & 119	& 5.0 \\
	\#3, 55107	($\hexagon$) & $1.5$	& $1.0$	& 113   & 5.5 \\
	\#4, 60751	($\bigtriangledown$) & $0.5$	& $0.5$	& 48	& 9.4 \\
	\#5, 79351	($\diamondsuit$) & $1.2$	& $0.6$	& 92	& 4.6 \\
	\#6, 92272	($\Diamond$) & $0.5$	& $0.3$	& 70	& 2.5 \\
	\#7, 99304	($\Box$) & $0.6$	& $0.4$	& 50	& 3.3 \\
	\hline
	\#8, 44316 & 0.9 & 0.7 & 309 & 0.9 \\
	\#9, 50682 & 1.4 & 0.3 & 45 & 2.8 \\
	\#10, 59076 & 0.9 & 0.3 & 66 & 1.6 \\
	\#11, 74682 & 0.5 & 0.5 & 114 & 2.4 \\
	\#12, 101499 & 0.9 & 0.2 & 65 & 0.5 \\
\end{tabular}
\end{table}

Figure~\ref{fig:selection} compares stellar mass, SFR, half-light radius, and gas-to-stellar-mass ratio of our parent sample of simulated galaxies to the observational $z\geq1.5$, $M_* > 4\times10^{10} M_{\odot}$ sample by \cite{Genzel20}.
Here, and for the remainder of this paper, we quote SFRs and masses of the TNG50 galaxies within a three-dimensional aperture with radius 20~kpc around the potential minimum, which corresponds to the size of our mock data cubes (see Section~\ref{s:mock}). The half-light sizes quoted for TNG50 refer to the radius containing half of the three-dimensional rest-frame $V-$band luminosity.\footnote{
Projected two-dimensional sizes are typically lower by $5-40$ per cent, depending on the projection angle.}
The $V-$band luminosity is based on stellar population synthesis models by \cite{Bruzual03} after accounting for age, mass, and metallicity of the simulated stellar particles as described by \cite{Vogelsberger13}. For a galaxy at $z=2$, this approximates well {\it Hubble Space Telescope} F160 bandpass imagery, providing independent estimates of half-light sizes for the majority of the sample by \cite{Genzel20}.
The individual galaxies are shown on top of the underlying galaxy population at $1.5<z<2.5$ based on the 3D-HST catalogue \citep{Brammer12, Skelton14, Momcheva16} with selection cuts  log$(M_*/M_{\odot})>9.2$, $K_{\rm AB}<23$~mag, and for the middle panel also SFR$/M_*>0.7/t_{\rm Hubble}$. TNG50 galaxies at $z=2$ with $M_* > 4\times10^{10} M_{\odot}$ and SFR~$>50M_{\odot}$~yr$^{-1}$ but excluded from the main kinematic analysis are shown as crosses. In the left and middle panels, the TNG50 galaxies are normalized to the same observationally constrained star-forming main sequence and mass-size relations as are the observations.

As mentioned above, the selected G20 galaxies lie on average above the main sequence ($\Delta$MS~$\sim0.2$~dex). Due to the generally lower SFRs of simulated galaxies, most of the selected TNG50 galaxies lie below the observationally calibrated main sequence. However, compared to the TNG-calibrated main sequence, the selected TNG50 galaxies have as well a positive offset of $\Delta$MS$_{\rm TNG}\sim0.3$~dex (cf.\, dotted line in the left panel of Figure~\ref{fig:selection}).

The half-light sizes of the seven TNG50 galaxies included in the kinematic analysis are comparable to the selected G20 galaxies but somewhat smaller (middle panel). Similar to systematic differences in SFR when comparing to observations, differences in sizes of SFGs are known for the TNG model: different measures of half-light or half-mass sizes for simulated $M_* \sim10^{11}~M_{\odot}$ $z=2$ SFGs give sizes that are on average lower by factors of $1.5-2$ compared to observations \citep{Genel18, Pillepich19}.

Also the simulation-intrinsic total gas-to-stellar-mass ratios (within a 20~kpc radius, filled colored symbols and large crosses) are on average lower compared to scaling relation-based estimates for the G20 galaxies and the underlying galaxy population (right panel). Furthermore, this difference is likely underestimated since the gas mass scaling relations by \cite{Tacconi18} provide {\it molecular} gas-to-stellar-mass ratios, based on redshift, stellar mass, and main sequence offset of a galaxy. Therefore, estimates based on these scaling relations may correspond to lower limits on the total gas mass. 
However, if we use these scaling relations to estimate the expected molecular gas-to-stellar-mass ratios also for the selected TNG50 galaxies, we find somewhat better agreement with the data (open colored symbols and small crosses).

In Table~\ref{tab:tngprop} we list the physical properties of the selected (top seven rows) and excluded TNG50 galaxies (bottom five rows). 
Examples of $z=2$ TNG50 galaxies with different stellar mass and/or SFR properties are shown in Figure 11 by \cite{Pillepich19}.

\subsection{Mock observations}
\label{s:mock}

For each selected simulated galaxy we generate mock observations for five lines of sight, for a total of 35 mocks. We first align the coordinate system of the galaxy using its moment of inertia tensor of the star-forming gas, such that the galactic plane coincides with the $xy-$plane, and the axis of rotation with the $z-$axis. 
We then define a line of sight by an inclination angle with respect to the $z-$axis and an orientation angle with respect to the $x-$axis. The five lines of sight are equally spaced around the galaxy and correspond to the same inclination ($i=60^{\circ}$) and position angle (PA$_{\rm kin}=90^{\circ}$), which we keep fixed for our main analysis. This choice allows us to examine the rotation symmetry of the simulated galaxies using one-dimensional kinematic extractions.

For each line of sight we bin the star-forming gas cells into a cube in position-position-velocity space which we subsequently convert to angular size and wavelength, such that our final cube sampling is $0.05^{\prime\prime}\times0.05^{\prime\prime}\times2.45$~{\AA}. At $z=2$, $1^{\prime\prime}$ corresponds to $\sim8$~kpc, and $1$~{\AA} corresponds to $15$~km~s$^{-1}$ in $K-$band. 
The cube is centered spatially on the potential minimum and in velocity direction on the center-of-mass velocity of the stellar component of the galaxy (which differs insignificantly from that of the gas). Then it is convolved with a three-dimensional Gaussian with a FWHM of 2~kpc and $80~{\rm km~s^{-1}}$ in the spatial and velocity directions, respectively, to approximate the effects of the instrument point spread function (PSF) and line spread function (LSF) for instruments such as SINFONI at the VLT in adaptive optics mode.\footnote{
In reality, the PSF can be of more complex functional shape \citep[see][for SINFONI observations]{FS18}. In Appendix~\ref{ap:2comp}, we briefly discuss the impact of using a more complex PSF model on kinematic extractions and modelling results for galaxy \#3.}
The PSF and LSF are typically well known from observations of standard stars and sky lines.

We then convert the instantaneous SFR into H$\alpha$ luminosity \citep{Kennicutt98}. In Appendix~\ref{ap:dust}, we briefly discuss the effect of dust on the mock-observations and kinematic extractions. However, accounting for dust does not affect our main conclusions, and we therefore do not include it in the main part of our analysis.
To account for realistic noise properties, including from random and systematic sources, and in particular stemming from the strong night sky line emission in the near-IR, we embed the mock data cube into a real noise cube from a deep SINFONI observation at $z\approx2$ \citep[cf.][]{Genel12a}.
To avoid biases due to a specific realization, we also randomize the noise cube for each mock observation. In addition, the mock line emission is scaled to reproduce the typical signal-to-noise ratio ($S/N$) of deep high$-z$ observations, with an average $S/N$ per spaxel of $S/N\gtrsim20$ in the central regions.

\subsection{Kinematic extractions}
\label{s:extractions}

With our mock data cubes in hand, we derive the kinematic properties of the simulated galaxies following the same methods used by \cite{Genzel17, Genzel20}. 
First, we derive the two-dimensional projected H$\alpha$ velocity and velocity dispersion fields using {\sc{linefit}} \citep{Davies09, Davies11, FS09}. This code takes into account the instrument LSF and fits a Gaussian profile to the line spectrum of each spaxel of the data cube. 

For the extraction of one-dimensional kinematic profiles (the rotation curve and the dispersion profile), we go back to the mock data cube and place a pseudo-slit of width $0.24^{\prime\prime}$ ($2$~kpc) on the kinematic major axis of the galaxies to generate a position-velocity diagram. Through cuts in velocity direction of width 4 pixels ($\sim1.7$~kpc) we then extract one-dimensional line profiles for different positions along the kinematic major axis. From those, we extract the velocity and velocity dispersion as a function of distance from the center using {\sc{linefit}}. 

Through visual inspection, we exclude radial bins where the assumption of a Gaussian profile is not justified by the line shape. This primarily affects extractions close to the galactic centers, where a broad range of velocities is blended through the finite spatial resolution and beam-smearing, and some regions in the galaxy outskirts that are strongly affected by skylines.

We adjust the centers of the one-dimensional profiles based on the steepest velocity gradient along the kinematic major axis, and the peak of the dispersion profile. The so determined kinematic centers deviate from the mock cubes centers (see Section~\ref{s:mock}) by less then the PSF and LSF FWHM. 

Following \cite{Genzel17}, we assign minimum uncertainties of $\pm5$~km~s$^{-1}$ for the velocity and $\pm10$~km~s$^{-1}$ for the velocity dispersion to more realistically account for systematics when the formal fitting uncertainties become very small.

\subsection{Dynamical modeling}
\label{s:modeling}

For the modeling of our mock galaxies we use the dynamical fitting code {\sc{dysmal}} (\citealp{Cresci09, Davies11, WuytsS16, Uebler18}; S.~Price et al.\ in prep.).
There exist several techniques and codes for the dynamical modeling of galaxy kinematics, for instance GalPaK$^{\rm 3D}$ \citep{Bouche15}, {$^{\rm 3D}$\sc{barolo}} \citep{diTeodoro15}, and others \citep[e.g.][]{Jozsa07, Sellwood15, SylosLabini19}, including non-axisymmetric modeling. 
For consistency with the analysis by \cite{Genzel17, Genzel20}, we follow their methodology. 

{\sc{Dysmal}} is a forward-modeling code that allows for the combination of multiple mass components. It accounts for flattened spheroidal potentials \citep{Noordermeer08}, includes the effects of pressure support on the rotation velocity \citep{Burkert10}, accounts for beam-smearing effects through convolution with the two-dimensional PSF, and for the instrument LSF. Here, we use again a three-dimensional Gaussian with a FWHM of 2 kpc and 80 km~s$^{-1}$ in the spatial and velocity directions, respectively.
We assume a velocity dispersion $\sigma_0$ that is isotropic and constant throughout the disc.

Out of the different mass components and accounting for the mentioned physical and `instrumental' effects, {\sc{dysmal}} first creates a three-dimensional mass model which is then converted to position-position-velocity space to resemble a data cube. Following \cite{Genzel17, Genzel20}, we perform the regression between the model and the (mock) data by use of extracted one-dimensional velocity and dispersion profiles. To this end, kinematic profiles with the exact same specifications (slit width, cuts in velocity direction) as for the mock data cubes (Section~\ref{s:extractions}) are extracted from the three-dimensional model cube, at each step of the fitting procedure (Section~\ref{s:mcmc}). 

In the following paragraphs, we describe our dynamical model with its free and fixed parameters, and we summarize our model parameters in Table~\ref{tab:fitting}.

\subsubsection{Baryonic parameters}

We assume that the H$\alpha$ kinematics trace the underlying mass distribution. Systematic studies of representative high$-z$ SFGs have shown that on average sizes based on H$\alpha$ tracing the young star-forming regions are larger by a factor 1.1-1.2 compared to sizes based on stellar light \citep{NelsonE16b, Wilman20}. \cite{WuytsS16} have shown that adopting a larger effective radius would typically increase the total dynamical mass by about 0.06-0.08 dex \citep[see also][]{Uebler19}, but that it would not significantly alter the mass within the effective radius.
On the other hand, far-infrared observations have started to reveal important dust aggregations that may obscure the stellar light in the central regions of massive SFGs at $z\sim2$, suggesting that sizes based on stellar light might be overestimated \citep[e.g.][]{Tadaki17b, Tadaki20}. In Appendix~\ref{ap:dust}, we show as an example for one galaxy the effect of dust obscuration: the $S/N$ in the central regions is particularly affected, but the extracted kinematics beyond the inner $2-3$~kpc do not change.

We consider two baryonic components for the dynamical modeling: a thick exponential disc and a central bulge. For the disc, we adopt a ratio of scale height to scale length $h_z/R_d=0.2$, motivated by the observed fall-off in the distribution of axis ratios of SFGs in this stellar mass and redshift range \citep{vdWel14b}, and a S\'{e}rsic index $n_{S, \rm disc}=1$. Following \cite{Genzel17, Genzel20}, we assume for the bulge an effective radius $R_{e, {\rm bulge}}=1$~kpc and a S\'{e}rsic index $n_{S, {\rm bulge}}=4$ \citep{Lang14, Tacchella15a}.\footnote{
We note that the typical ratio of half-light height to half-light size for massive $z=2$ SFGs in TNG50 is closer to $h_{1/2}/R_{1/2}\sim0.1$ \citep{Pillepich19}, and from bulge-to-disc decompositions to the azimuthally averaged baryon distribution we find bulge sizes $R_{e, {\rm bulge}}<1$~kpc for our sample. However, because these quantities are typically hard to measure observationally for individual, high$-z$ galaxies, we proceed with these typically adopted values for modeling of observational data.}

For the baryonic disc effective radius $R_e$, we use a Gaussian prior centered on the rest-frame $V-$band half-light radius $R_{1/2}$, with a standard deviation of 1~kpc and hard bounds of 1.5-12~kpc. Using $R_{1/2}$ as a prior provides an initial (although somewhat uncertain, see discussion above) guess for the disc size that is in principle expected to be larger for bulgy galaxies. 
For the total baryonic mass we use a Gaussian prior centered on the intrinsic value with a standard deviation and hard bounds of 0.2~dex and $\pm0.5$~dex. 
With this approach, we fold into our modeling the typical uncertainties on those parameters expected from observational data.
We assume a flat prior for $B/T$ between 0 and 0.6, motivated by the typical values expected for SFGs in this stellar mass range and redshift \citep[see][]{Lang14}. 
For the intrinsic velocity dispersion $\sigma_0$, we adopt a flat prior between 10 and 100~km s$^{-1}$, covering the range of values observed in SFGs at $z\sim2$ \citep[see][and references therein]{Uebler19}.

As mentioned in Section~\ref{s:mock}, we fix the disc inclination and position angle to their true values, as defined by the moment of inertia tensor of the star-forming gas. In doing so, we put our focus on any kinematic asymmetries, and their impact on the dynamical modeling results. 
For the observations by \cite{Genzel17, Genzel20}, the inclination is inferred from the minor-to-major axis ratio of the stellar light distribution, known from ancillary data, and fixed for high-inclination systems. Changes/uncertainties in inclination translate directly into changes/uncertainties in the dynamical mass estimate \citep[see e.g.][]{Uebler18}. 
Therefore, we have tested including the inclination as a model parameter in the range $i=30-90^{\circ}$. The true inclination of $i=60^{\circ}$ is typically recovered within the 1$\sigma$ MCMC posterior distribution, and the effect on the other model parameters is minor.

\subsubsection{Dark matter density profile}\label{s:dmdensity}

\begin{figure}
	\centering
	\includegraphics[width=\columnwidth]{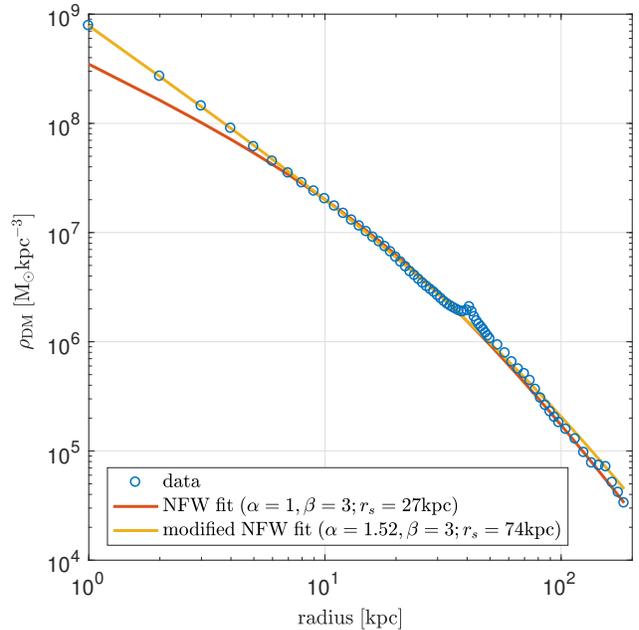}
	\caption[Example of two-power density fit to the dark matter halo]{Two-power density fit to the spherically averaged dark matter density distribution of the halo of galaxy \#3 from TNG50. On the galaxy scale ($r\lesssim20$~kpc) and beyond, the dark matter density (blue dots) is well fit by a modified NFW halo with $\alpha=1.52$ (yellow line), while a standard NFW fit with $\alpha=1$ (red line) underestimates the central dark matter density. 
	This halo has a virial radius $R_{200,c}\approx174$~kpc with a total mass of $\log(M_{200,c}/M_{\odot})=12.7$. The substructure visible at a distance of $40$~kpc hosts a companion galaxy with a stellar mass ratio to the central galaxy of 1:5 (see Section~\ref{s:intkin}).
	For our modified NFW fit we find a scale radius $r_s\approx74$~kpc, corresponding to a concentration parameter of $c=2.4$. For the pure NFW fit we find $r_s\approx27$~kpc, corresponding to $c=6.4$.}
	\label{f:dmdensity}
\end{figure}

The results by \cite{Genzel20} that we compare to in this work assume a standard NFW dark matter halo profile for the dynamical modeling, with adiabatic contraction effective, or not (but see their modeling results with free $\alpha$; Eq.~\ref{eq:twopower}). Through modified fits to the intrinsic dark matter density distributions of the TNG50 galaxies with $\beta=3$ but $\alpha$ as a free parameter, we find that all simulated haloes have a steeper inner slope with respect to a pure NFW halo, with individual values of $\alpha=1.4-1.7$. These values indicate contractions of the dark matter haloes \citep[see also][]{Lovell18}. An example is shown in Figure~\ref{f:dmdensity}: the intrinsic halo profile (blue circles) is well-fitted by a modified NFW halo with $\alpha=1.5$ (yellow line; except for the companion at $r\sim40$~kpc), whereas a pure NFW fit underestimates the dark matter density on galactic scales (red line). 

Through these fits to the intrinsic dark matter density distribution, we also constrain the halo concentration parameter $c=R_{200,c}/r_s$. The scale radius, $r_s$, is defined to be the radius where the slope of the density profile equals $-(\beta+\alpha)/2$. This is by definition $-2$ for an NFW halo, but varies for our modified NFW haloes with values $\leq-2$, leading to larger scale radii (see also Figure~\ref{f:dmdensity}).

For our fiducial models, we adopt both unmodified and modified NFW profiles for the dark matter distribution, and leave the total halo mass as a free parameter between $M_{200,c}=10^{11}-10^{13.5} M_{\odot}$. 

\begin{table}
\caption{Summary of free and fixed model parameters with their priors/values. "tG($x;y;z$)": truncated Gaussian prior with (center $x$; width $y$; bounds $z$), "F[$x;y$]": flat prior in range [$x;y$]. Values taken directly form the simulation or from fits to simulation intrinsic data are indicated as `true'.}
\label{tab:fitting}
\centering
\begin{tabular}{lc}
\hline
	free model parameter & prior \\
\hline
    baryonic mass log($M_{\rm bar}/M_{\odot}$) & tG(true; 0.2; $\pm$0.5) \\
    disc effective radius $R_e$ [kpc] & tG($R_{1/2}$; 1; [1.5;12]) \\
    baryonic bulge-to-total fraction $B/T$ & F[0; 0.6] \\
    intrinsic velocity dispersion $\sigma_0$ [km s$^{-1}$] & F[10; 100] \\
    dark matter mass log($M_{\rm halo}/M_{\odot}$) & F[11; 13.5] \\
\hline
	fixed model parameter & value \\
\hline
    disc S\'{e}rsic index $n_{S,\rm{disc}}$ & 1 \\
    disc thickness $h_z/R_d$ & 0.2 \\
    bulge effective radius $R_{e,\rm{bulge}}$ [kpc] & 1 \\
    bulge S\'{e}rsic index $n_{S,\rm{bulge}}$ & 4 \\
    inclination $i$ [$^{\circ}$] & true \\
    PA$_{\rm kin}$ [$^{\circ}$] & true \\
    halo concentration parameter $c$ & true \\
    modified NFW halo inner slope $\alpha$ & true \\
\end{tabular}
\end{table}

\subsubsection{MCMC setup}\label{s:mcmc}

Using {\sc{dysmal}}, we simultaneously fit the extracted one-dimensional velocity and velocity dispersion profiles. We apply Markov chain Monte Carlo (MCMC) sampling using the {\sc emcee} package by \cite{ForemanMackey13} to determine the model likelihood based on comparison to the extracted profiles, and assuming Gaussian measurement uncertainties. 
To ensure convergence of the MCMC chains, we model each galaxy with $\geq200$ walkers per free parameter, and a burn-in phase of 100 steps followed by a running phase of 200 steps ($>10$ times the maximum auto-correlation length of the individual parameters). For each free parameter, we adopt the median of all model realizations as our best fit value, with symmetric uncertainties corresponding to half the difference between the $16^{\rm th}$ and $84^{\rm th}$ percentile of the one-dimensional marginalized posterior distributions.

\section{Results}\label{s:results}

\subsection{Intrinsic kinematics}\label{s:intkin}

\begin{figure*}
	\centering
	\includegraphics[width=0.8\textwidth]{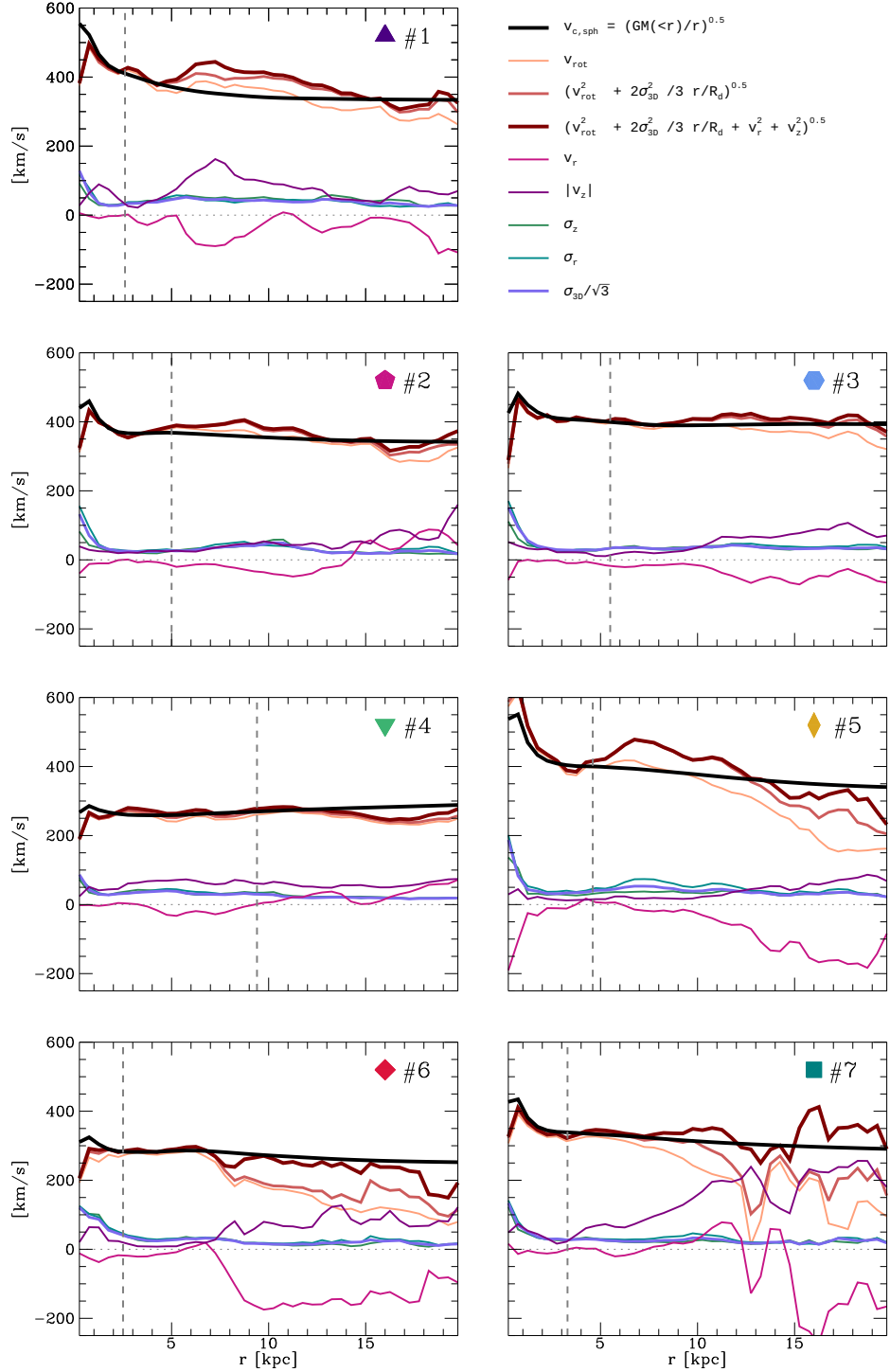}
	\caption[Intrinsic one-dimensional kinematics]{Different measures of the intrinsic velocity and velocity dispersion of a selection of TNG50 simulated galaxies, as indicated in the legend. The circular velocity $v_{\rm c,sph}$ (black line) is calculated from the enclosed mass assuming spherical symmetry. The radial velocity dispersion $\sigma_r$ (turquoise), vertical velocity dispersion $\sigma_z$ (green), and three-dimensional velocity dispersion $\sigma_{\rm 3D}/\sqrt{3}$ (blue) are measured `locally' in $xy$ bins of 0.5~kpc length, and subsequently averaged, as is the vertical velocity component $v_z$ (purple lines). All other properties are azimuthal averages. Light and dark brown lines show different corrections to the rotation velocity $v_{\rm rot}$ (salmon) from turbulent and other non-circular motions. Grey vertical dashed lines indicate the location of the stellar rest-frame $V-$band half-light radii (see Table~\ref{tab:tngprop}). Beyond $r\sim2$~kpc, the velocity dispersion is approximately constant with radius. All galaxies show substantial radial (magenta) and vertical motions.}
	\label{f:intrvel}
\end{figure*}

Before we present the results of our mock observations and modeling, we first discuss intrinsic kinematic properties of the TNG50 sample to create a basis for our further discussion.
Not all details simulated are accessible for real galaxies, but their study can highlight effects that are potentially relevant to observational work.

In Figure~\ref{f:intrvel} we show different measures of the intrinsic, one-dimensional velocity and velocity dispersion profiles for all selected galaxies. 
The radial velocity dispersion $\sigma_r$ (turquoise), the vertical velocity dispersion $\sigma_z$ (green), and the three-dimensional velocity dispersion $\sigma_{\rm 3D}/\sqrt{3}$ (blue) are measured `locally', i.e.\ in $xy$ bins of 0.5~kpc length, and subsequently averaged, as is the vertical velocity component $v_z$ (purple lines show $|v_z|$). All other properties are azimuthal averages: the rotation velocity $v_{\rm rot}$ (salmon), radial velocity $v_r$ (magenta), and modifications to $v_{\rm rot}$ including velocity dispersion (light brown) and vertical and radial motions (dark brown). 
For the velocity dispersion measures, the final azimuthal averages are luminosity-weighted.
In addition, we show as a reference the circular velocity $v_{\rm c,sph}(r)=\sqrt{G\cdot M(<r)/r}$, calculated from the enclosed mass under the assumption of a spherically symmetric potential (black line). Note that the rotation curve of a thick exponential disc has a peak velocity that is about 10 per cent higher compared to a spherical distribution of the same mass \citep[e.g.][]{Casertano83, BT08}.

The three different measures of the intrinsic velocity dispersion agree well, suggesting that the velocity dispersion is fairly isotropic. Furthermore, beyond $r\sim1-2$~kpc the velocity dispersion is remarkably constant, suggesting the existence of a galaxy-wide pressure floor,\footnote{
Most star-forming gas in the simulation has effective temperatures smaller than a few $10^4$~K \citep[see][]{Springel03}, corresponding to sound speeds lower than or similar to the velocity dispersions we find. In Section~\ref{s:veldisp} we briefly discuss the effect of including a `thermal term' \citep{Pillepich19} to account for unresolved or sub-grid gas motions.} 
consistent with other galaxy formation models \citep[e.g.][]{Teklu18, Wellons19}. 
High-quality, adaptive-optics resolution observations of individual galaxies support roughly constant and isotropic velocity dispersions \citep{Genzel11, Uebler19}. 
At the half-light radius (vertical grey dashed line), the velocity dispersion measures are typically below 50~km~s$^{-1}$.

Out to $r\sim10$~kpc and sometimes beyond, the gas rotation velocity $v_{\rm rot}$ approximately traces $v_{\rm c,sph}$. The light brown lines show the rotation velocity corrected for pressure support following \cite{Burkert10} and using the three-dimensional velocity dispersion $\sigma_{\rm 3D}(r)/\sqrt{3}$. Due to the high rotation velocities ($\sim250-450$~km~s$^{-1}$) and the moderate velocity dispersion, the relative effect of this pressure support correction is small for most simulated galaxies. 
For galaxies \#1, \#2, and \#5, the pressure-corrected $v_{\rm rot}$ overshoots $v_{\rm c,sph}$ at $r\sim5-10$~kpc. Due to the different assumption about the mass distribution for $v_{\rm c,sph}$, this is not unexpected.

All simulated galaxies show substantial amounts of vertical $v_z$ and radial motions $v_r$. The magnitudes of these motions are often correlated (e.g., galaxy \#1, top left panel), suggesting streaming motions diagonal to the galactic plane, possibly related to minor mergers (as opposed to pure radial inflow triggered by bar or disc instabilities). To assess the impact of these additional non-circular motions, we `correct' the rotation velocity in a simplistic attempt not only for pressure support, but also for radial and vertical motions as follows:\footnote{
In Appendix~\ref{ap:asym}, we briefly describe the effect of vertical and radial motions on the kinematic extraction of rotation velocity and velocity dispersion.
}
\begin{equation}\label{eq:vcorr}
    v_{\rm circ}(r)\equiv \sqrt{v_{\rm rot}^2(r)+2\frac{\sigma_{\rm 3D}^2(r)}{3}\frac{r}{R_d} + v_r^2(r) + v_z^2(r)}
\end{equation}
This correction leads to the dark brown line, which in some cases corresponds better to $v_{\rm c,sph}$ \citep[see also][]{Wellons19}, seen especially in the outer regions of galaxies \#5-7 at $r\gtrsim7$~kpc (beyond their visible extent). These motions, however, most likely correspond to low-surface brightness, misaligned accreting gas. 

In general, the non-circular motions present in the TNG50 galaxies could also explain the sometimes large differences in rotation curve shapes at fixed inclination that we discuss later on in Section~\ref{s:asym} \citep[see also][]{Oman19}.\\

{\bf Comments on environment: }
Some of the deviations from circular motions in the simulated galaxies could stem from high gas accretion rates or tidal interactions with other massive galaxies (not in our main sample), factors not captured by our initial selection criteria.
This is plausible inasmuch as the selected simulated galaxies have a positive offset with respect to the {\it TNG50 main sequence} (see discussion in Section~\ref{s:selection}), possibly corresponding to more irregular kinematics due to relatively higher accretion rates or merging.
Indeed, galaxies \#2, \#3, and \#4 have in their vicinity ($\Delta r=30-60$~kpc) another massive galaxy (mass ratio $\geq$1:2 for \#2 and \#4, and mass ratio 1:5 for \#3). 
Galaxies \#2, \#3, and \#4 also show asymmetries in their extracted rotation curves, and among different lines of sight (see Figures~\ref{f:1dlos} and \ref{f:1dloscomp}).
Galaxies \#2 and \#4 will merge with their companion within the next 1~Gyr, and galaxy \#4 has also just completed a 1:5 merger $\sim100$~Myr ago.
Galaxy \#1 has a companion with mass ratio $\geq$1:2, but at a distance of 140~kpc. 
In contrast, galaxies \#5, \#6, and \#7 are sufficiently isolated from similarly massive objects to be considered undisturbed, supporting the above interpretation of accreting gas being responsible for the large vertical and radial motions beyond $r=7$~kpc.

Five galaxies in the selected G20 sample have potentially close ($\Delta r=6-21$~kpc) but low-mass (mass ratios 1:8$-$1:50) companions \citep[see Table 1 by][]{Genzel20} --  only one of the former is with $\Delta$MS~$\approx0.55$ substantially above the main sequence. As discussed in detail by \cite{Genzel17, Genzel20}, even such smaller satellites can in theory affect the kinematics of the main galaxy, if they are close enough.
In Section~\ref{s:asym} we present an analysis of the symmetry properties of the simulated and observed rotation curves. From this analysis, we do not see any evidence that these five systems are systematically more asymmetric compared to the other galaxies without companions. We refer the reader to \cite{Genzel20} for a more in-depth presentation of the environmental properties of the full observational sample.

\begin{figure*}
	\centering
	\includegraphics[width=\textwidth]{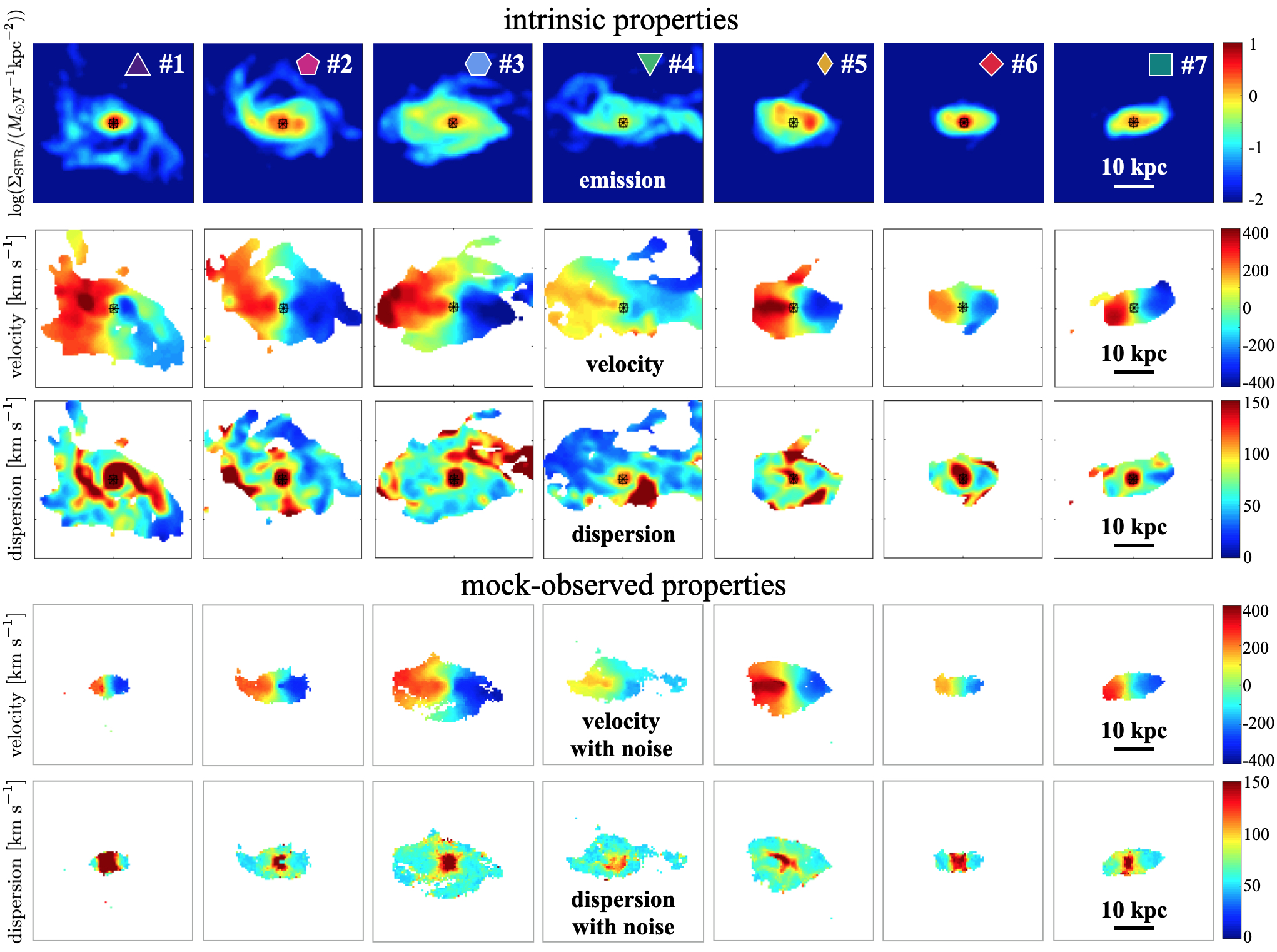}
	\caption[Projected two-dimensional kinematic maps]{Projected two-dimensional maps of the star-forming gas and its kinematic properties of the seven TNG50 selected galaxies. 
	The top three rows show PSF-convolved intrinsic parameters (top row: $\Sigma_{\rm SFR}$; second row: velocity; third row: velocity dispersion). 
	The bottom two rows show the velocity and dispersion fields with $S/N\geq5$ after including realistic noise, discretization into pixels, and Gaussian line fitting (fourth row: velocity; bottom row: velocity dispersion) for the seven selected TNG50 galaxies (columns). The projections correspond to an inclination of $i=60^{\circ}$. The panels show 40~kpc~$\times$~40~kpc in projection.
	The mock velocity and dispersion fields retain a large amount of the information content of the intrinsic kinematics in regions of high star-formation rate surface density.}
	\label{f:2dmaps}
\end{figure*}

\begin{figure*}
	\centering
	\includegraphics[width=0.9\textwidth]{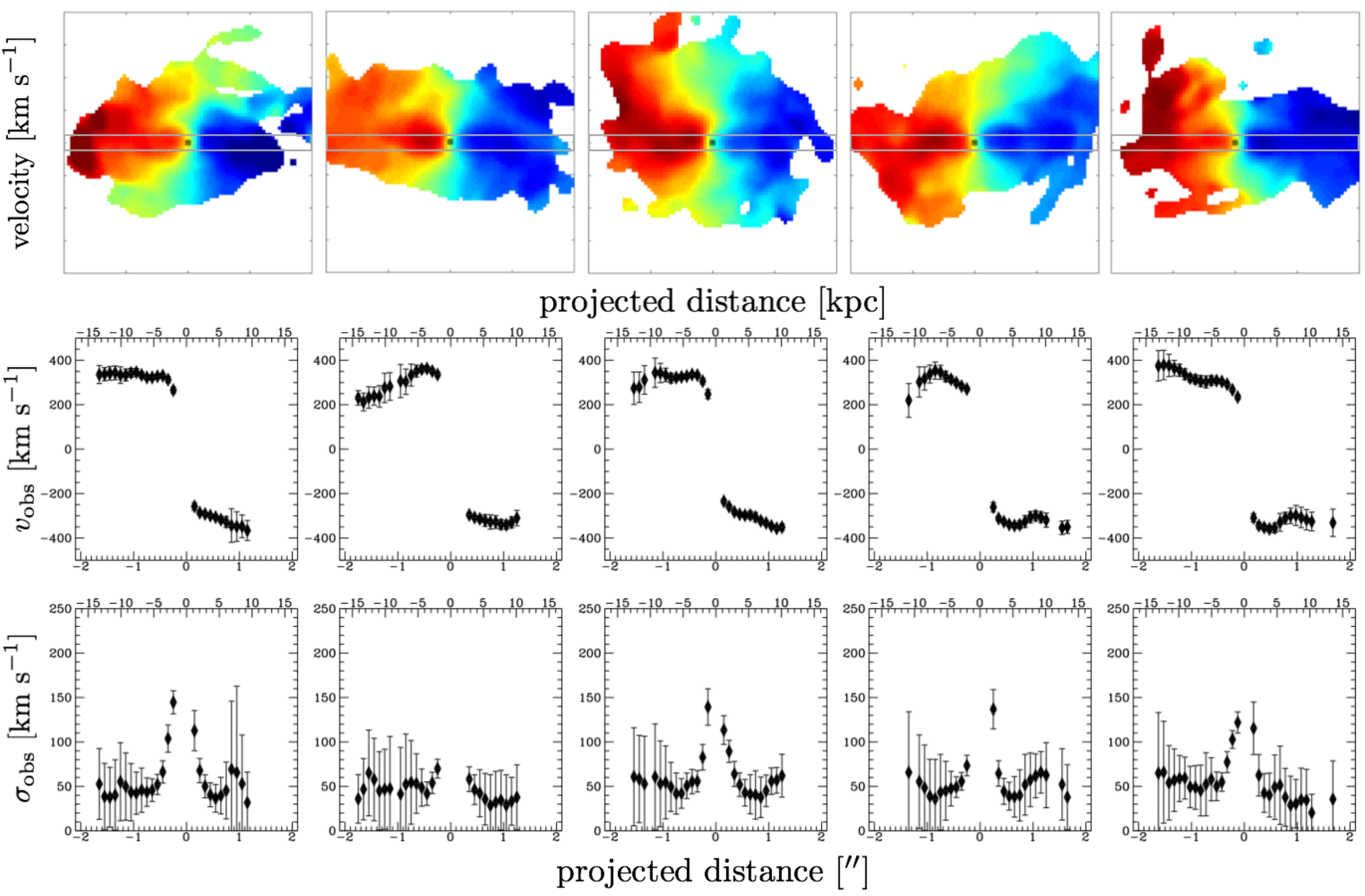}
	\caption[Extracted kinematics along different lines of sight I]{Extracted kinematics along five different lines of sight (columns) for galaxy \#3 from TNG50. The inclination is always $i=60^{\circ}$. Top row:  noise-free, convolved velocity map, 40~kpc~$\times$~40~kpc in projection, with colors corresponding to [-400; 400] km~s$^{-1}$; second row: extracted velocity along the kinematic major axis; bottom row: extracted velocity dispersion along the kinematic major axis, corrected for the `instrument' LSF. Otherwise, the extracted profiles are shown in `observed space', i.e.\ not corrected for inclination and `beam smearing'. 
	Not shown here are velocity and dispersion extractions for emission regions with highly non-Gaussian line profiles, typically found in the central $0.2-0.5^{\prime\prime}$ of the mock data. The horizontal grey lines in the top row panels illustrate the pseudo slit used for the one-dimensional kinematic extractions. 
	The kinematic extractions correspond well to the intrinsic kinematic features: 
	particularly the rotation velocities can differ substantially along different lines of sight. Variations in the velocity dispersion are more modest, and typically within the uncertainties, which can however become large in the outer regions.}
	\label{f:1dlos}
\end{figure*}

\subsection{Kinematic extractions}

We now turn to the results of our mock analysis. 
In Figure~\ref{f:2dmaps} we compare for one line of sight per galaxy the two-dimensional maps of noise-free $\Sigma_{\rm SFR}$, velocity, and velocity dispersion after convolution with the PSF (three top rows) with the velocity and velocity dispersion maps with $S/N\geq5$ after accounting for realistic noise, discretization into pixels, and Gaussian fitting with {\sc linefit} (two bottom rows). Through the addition of noise in the mock observations, the fainter emission in the outskirts of the galaxies is no longer visible, including low-surface brightness inflows and tidal features. 
However, above $S/N=5$ the mock velocity and dispersion fields well reproduce the intrinsic kinematics.

To investigate the regularity of the simulated kinematics, we extract for each galaxy two- and one-dimensional kinematics from the mock observations along five equally spaced lines of sight, but keeping the inclination fixed to $i=60^{\circ}$. 
In Figure~\ref{f:1dlos} we compare the one-dimensional velocity and velocity dispersion profiles extracted along the five different lines of sight for galaxy \#3, for which the high surface brightness region is most extended. 
As is the case for the two-dimensional maps, the one-dimensional extracted rotation curves qualitatively compare well to the intrinsic velocity fields, modulo noise. I.e., falling/rising rotation curves or wiggles are easily associated to corresponding velocity changes in the intrinsic velocity fields along the major axis.

However, the kinematics can vary substantially between different lines of sight, with differences in the outer rotation velocities of up to $\sim200$~km~s$^{-1}$, much larger than their typical uncertainties of $\sim30-50$~km~s$^{-1}$ in the outer regions. The velocity dispersions are generally more similar along different lines of sight, with maximum variations in the outer regions of $\sim30$~km~s$^{-1}$ and typical uncertainties on the extracted values of $\sim30-60$~km~s$^{-1}$.

\begin{figure*}
	\centering
	\includegraphics[width=0.3\textwidth]{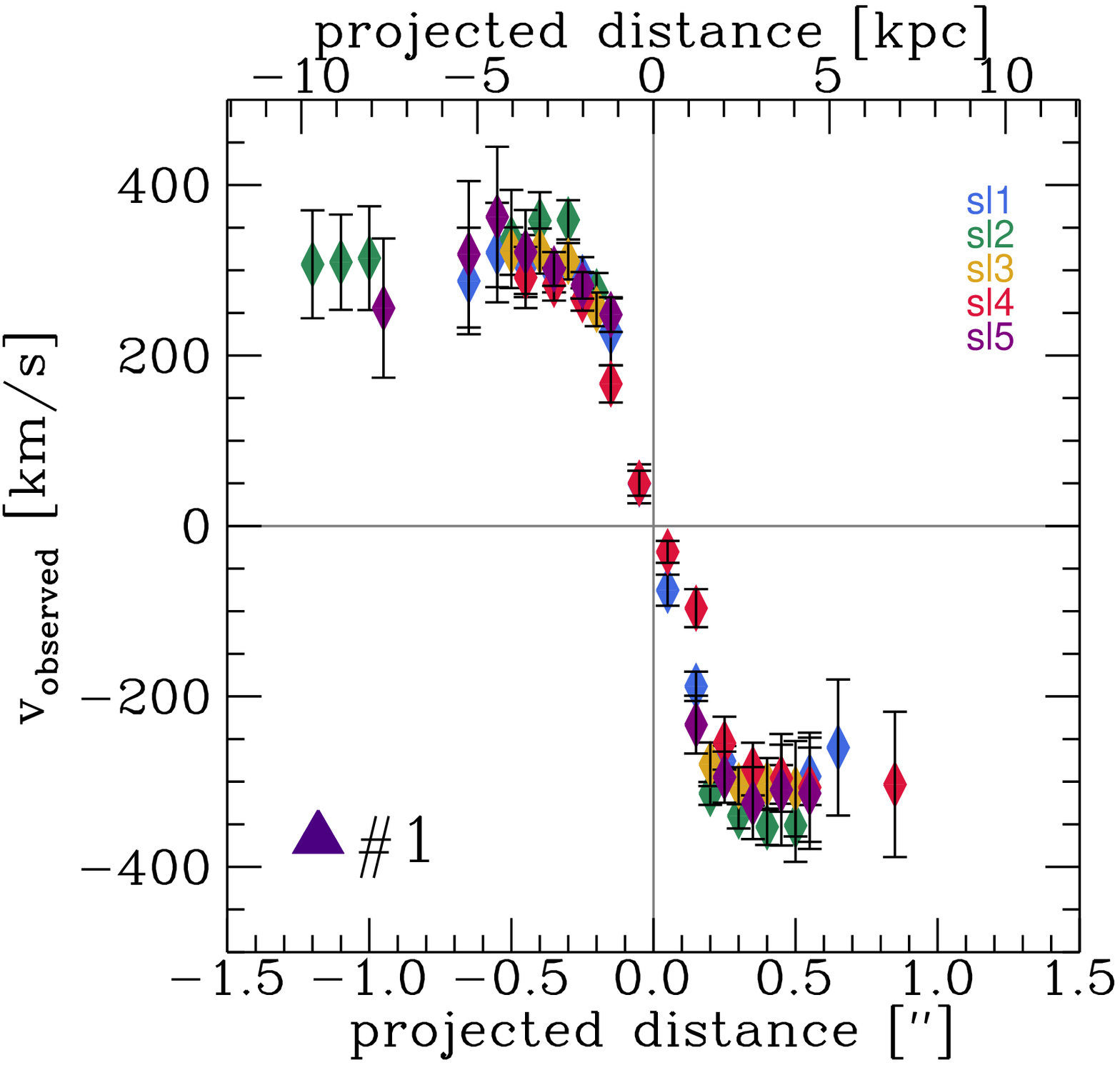}
	\includegraphics[width=0.3\textwidth]{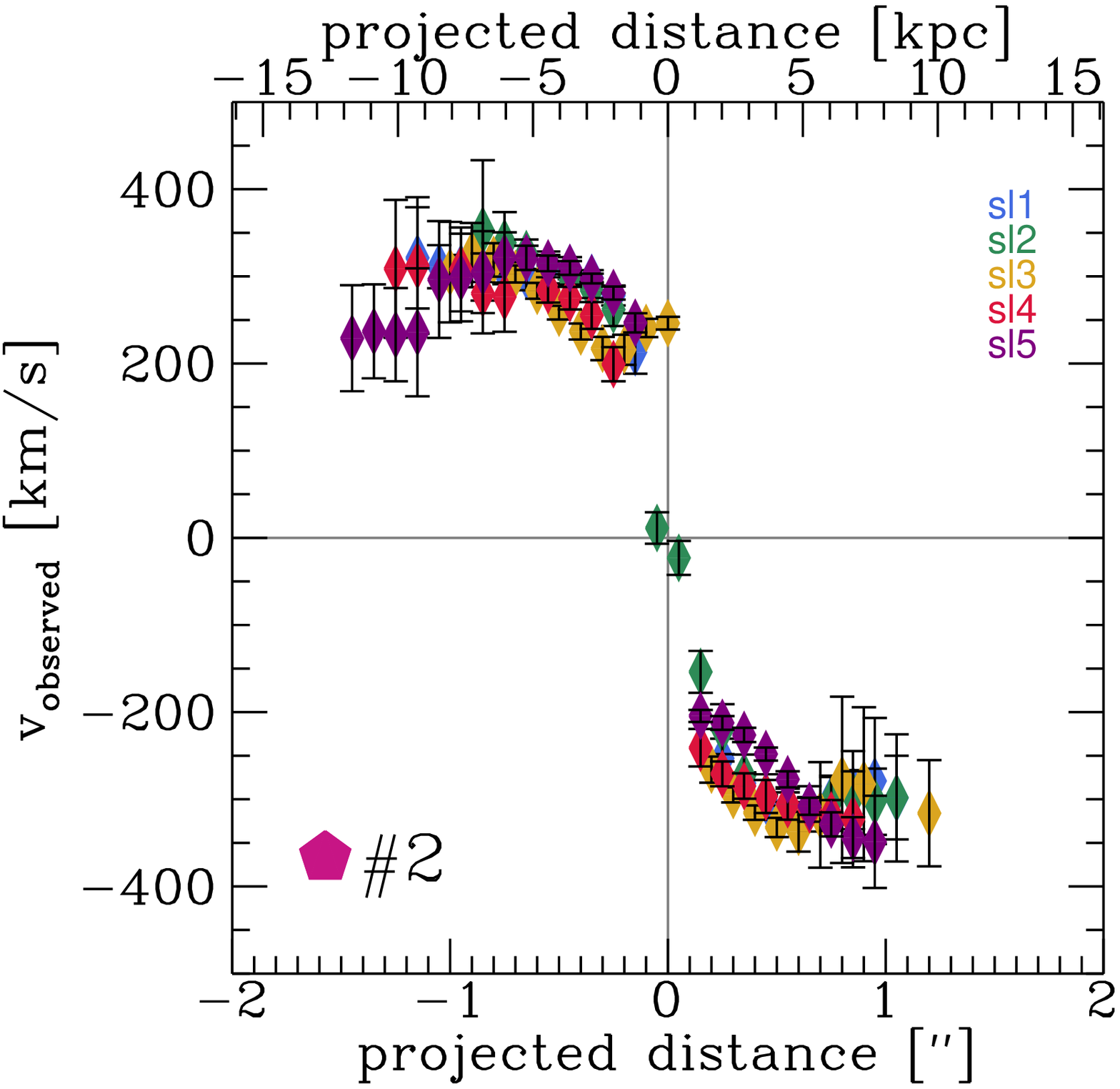}
	\includegraphics[width=0.3\textwidth]{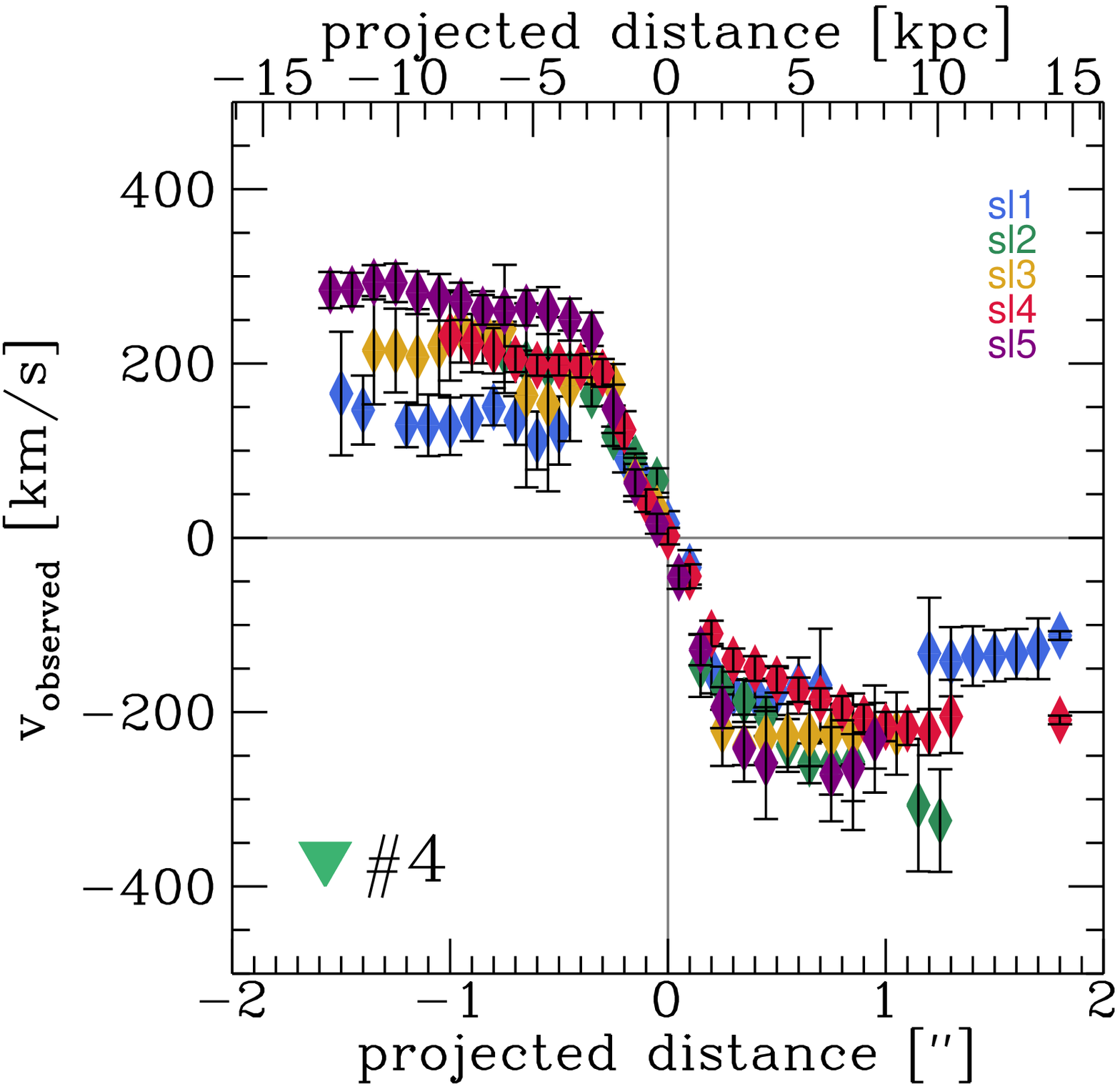}
	\includegraphics[width=0.3\textwidth]{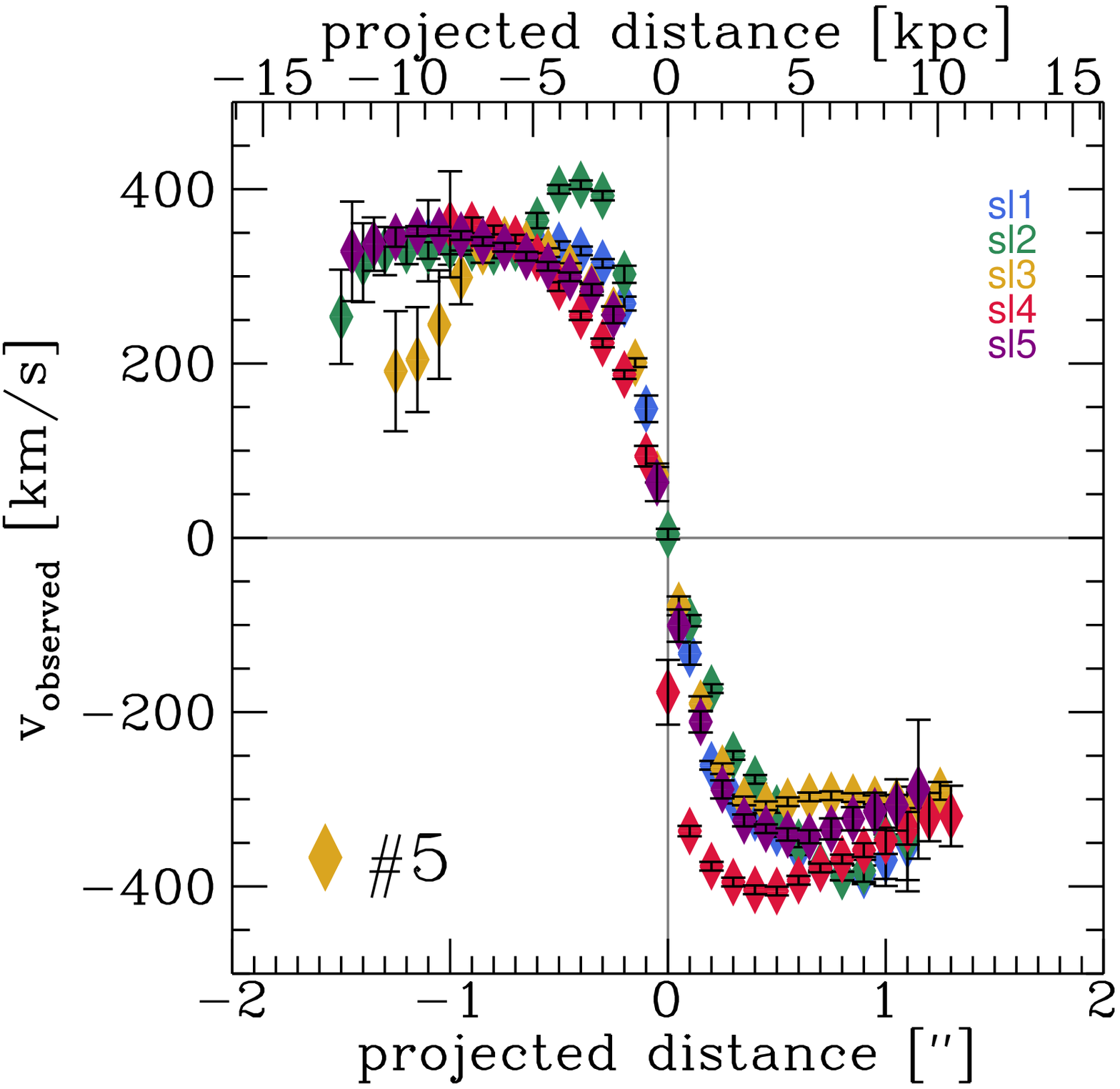}
	\includegraphics[width=0.3\textwidth]{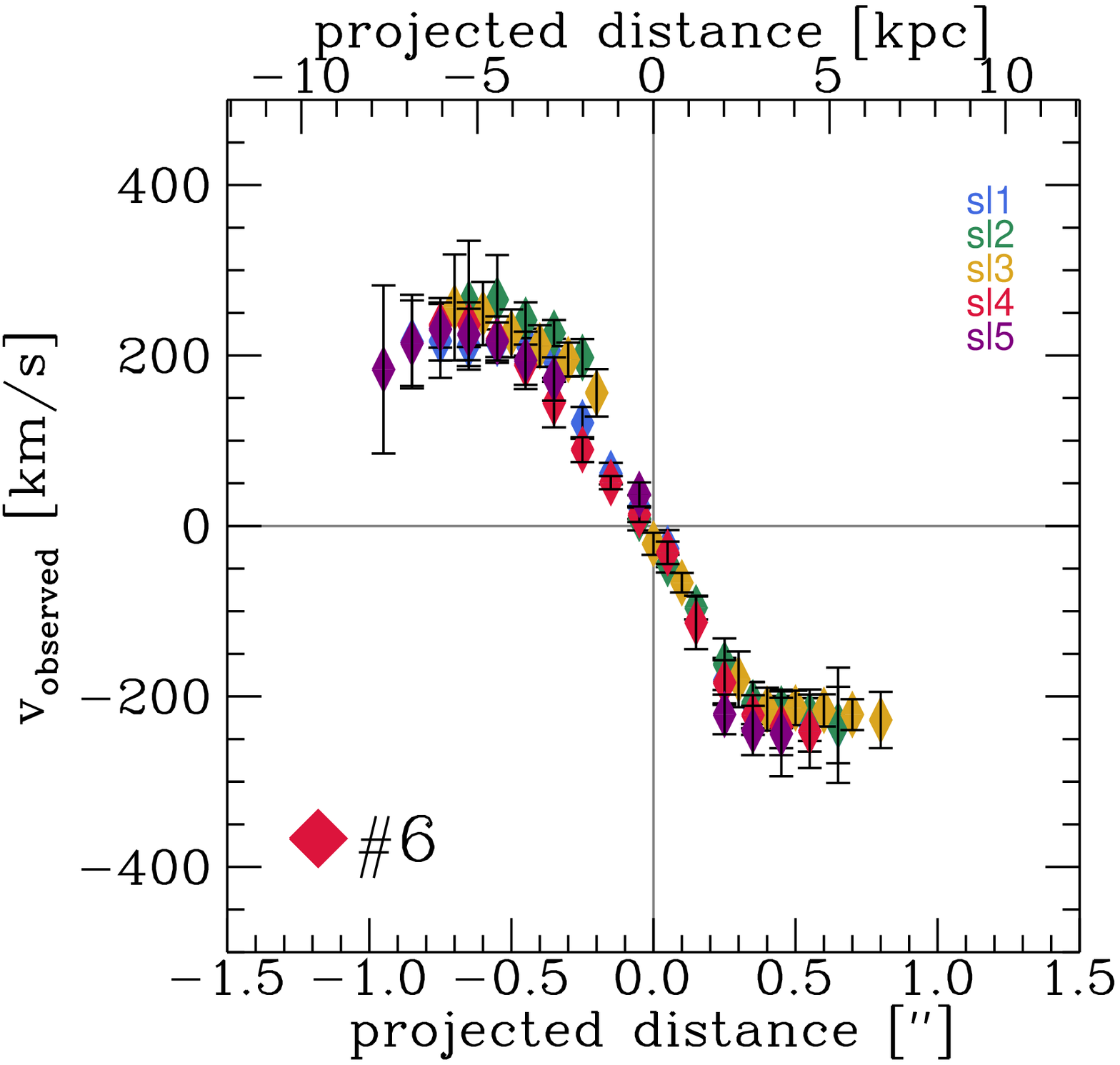}
	\includegraphics[width=0.3\textwidth]{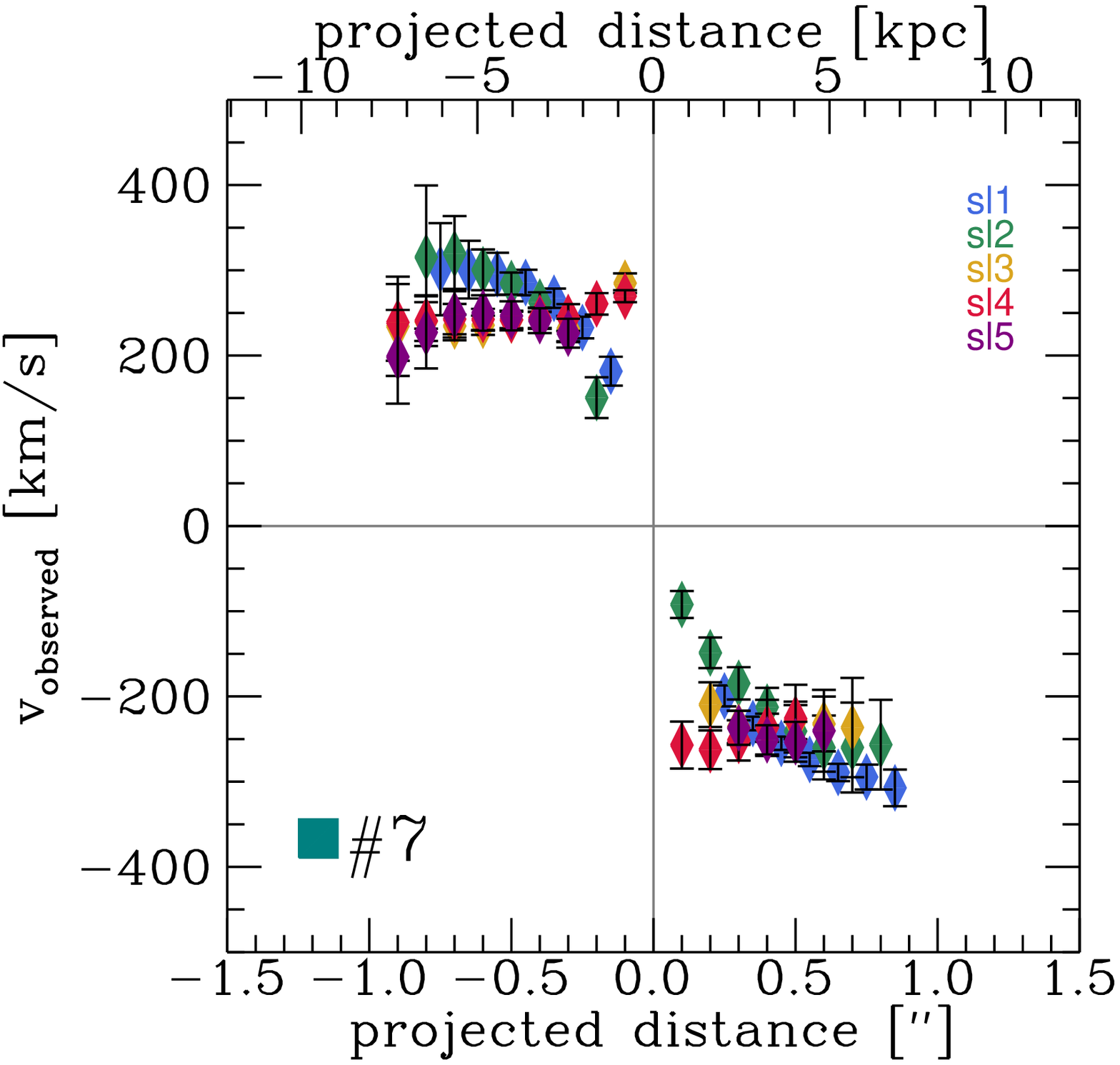}	
	\caption[Extracted kinematics along different lines of sight II]{Extracted velocity profiles along five different lines of sight (colors) for galaxies \#1, \#2, \#4, \#5, \#6, and \#7 (from left to right and top to bottom). The rotation curves are shown in `observed space', i.e.\ not corrected for inclination and `beam smearing'. We find variations in extracted major axis velocities along different lines of sight, particularly for galaxies \#4, \#5, \#7, and \#3 (Figure~\ref{f:1dlos}).}
	\label{f:1dloscomp}
\end{figure*}

For each line of sight, the extracted rotation curves show statistically significant asymmetries from their approaching to receding sides. These asymmetries are seen in the intrinsic data \citep[see Figure 11 by][for additional examples of intrinsic kinematics of $z=2$ TNG50 galaxies]{Pillepich19}. 
In Figure~\ref{f:1dloscomp} we show the major axis rotation velocities for the other six galaxies, where different colors indicate the five different lines of sight (all with the same inclination). Again, strong asymmetries for individual lines of sight as well as large differences between different lines of sight  are evident. The galaxies with the most symmetric individual lines of sight (\#6 and \#1) also have best-fit dynamical models with the highest success rate in recovering the intrinsic central dark matter fraction (see discussion in Section~\ref{s:recovery}).

\subsection{Kinematic asymmetries}\label{s:asym}

To quantify the asymmetry of the mock-observed rotation curves and compare it to that of real galaxies, we use two methods.
The first method employs the overlapping coefficient.
For each radial bin in a rotation curve we define two normal distributions centered on the absolute value of the receding and approaching velocities, with widths given by the velocity uncertainties. We then calculate the overlapping area of the normal distributions (a value between 0 and 1, where 0 indicates no overlap and 1 indicates complete overlap, including uncertainties). This is done for each radial bin, if necessary using linear interpolation, and then divided by the sum of radial bins to get a normalized value for each rotation curve. 
For the TNG50 galaxies, we find values of 0.09 to 0.73 (0.20 to 0.64 when averaged among different lines of sight per galaxy), with a median 0.46 (0.45).
In the comparison G20 sample, we find values of 0.38 to 0.81, with a median of 0.58.
Because this measure is directly affected by the signal-to-noise ratio, we recalculate the coefficient by fixing the width of the normal distributions (i.e. the velocity uncertainty) to a common value of 10~km~s$^{-1}$, to capture the rotation curve asymmetry independent of $S/N$. While the values for the selected G20 galaxies change slightly, for a median value of 0.57, for the simulated galaxies we now find a substantially lower median of 0.27. 

We use a second method to characterise asymmetry: we fit a quadratic function to one side of the rotation curve, calculate the reduced chi-squared statistics ($\chi^2_{\rm red}$), and compare it to the other side of the rotation curve through point reflection. The difference between the goodness-of-fit for both sides is $\Delta\chi^2_{\rm red}$, and we calculate it independently for fits to both sides of the rotation curve. We find average values of $\chi^2_{\rm red}\approx0.9$ when considering only one side of a rotation curve, indicating that small values of $\Delta\chi^2_{\rm red}\sim1$ would correspond to both good fits and symmetric rotation curves. We find mean (median) $\Delta\chi^2_{\rm red}=46.8$ (5.8) for the simulated galaxies, and mean (median) $\Delta\chi^2_{\rm red}=2.1$ (1.6) for the G20 sample. 
Again, we repeat our calculations for fixed velocity uncertainties of $\delta v_{\rm rot}=\pm10$~km~s$^{-1}$. For the TNG50 galaxies, we find mean (median) $\Delta\chi^2_{\rm red}=41.4$ (18.1), while for the G20 sample we find mean (median) $\Delta\chi^2_{\rm red}=3.7$ (3.4). This test shows that the large $\Delta\chi^2_{\rm red}$ we find for the TNG50 galaxies is not due to $S/N$, but is because the rotation curves are less symmetric.

For the TNG50 sample, the results from individual galaxies also agree well with the visual impression from Figure~\ref{f:1dloscomp}: the most symmetric rotation curves are extracted for galaxies \#6 and \#1, while the most asymmetric ones are extracted for galaxy \#5. 

Both methods demonstrate that the simulated galaxies show large asymmetries in their rotation curves, and are less regular compared to the galaxies observed by \cite{Genzel20}.\footnote{
Some galaxies in our G20 comparison sample have a larger FWHM ($\sim5$~kpc) due to seeing-limited observations. To test the effect of a larger PSF size on the (a)symmetry of our simulated sample, we repeated our mock-observation and kinematic extractions for galaxy \#3, now mimicking seeing-limited data. Perhaps surprisingly, we find no systematic effect of the PSF size on the overall symmetry of the extracted kinematics using both measures described above. This is probably because we trace kinematics out to large distances; for systems that are barely or not resolved, a systematic effect would likely be registered.}
For their full sample, \cite{Genzel20} find that only 3/41 rotation curves show significant deviations from reflection symmetry.
Certainly, asymmetric kinematics exist for high$-z$ SFGs. While we cannot make a general quantitative statement based on the relatively low number statistics of the simulated and observed sample, we conclude that the mass- and SFR-matched TNG50 galaxies are more asymmetric in their kinematics compared to the observational G20 comparison sample.

\subsection{Dynamical modeling performance}\label{s:recovery}

For all seven galaxies, we model the one-dimensional kinematics extracted from our mock observations along five lines of sight with {\sc dysmal}. Generally, asymmetric kinematics along the major axis hamper successful modeling with {\sc dysmal} because the code assumes axisymmetric mass distributions.
Using a non-axisymmetric modeling approach instead may help to facilitate the recovery of complex intrinsic galaxy kinematics. On the other hand, this would complicate the derivation of baryonic and dark matter mass fractions, quantities we are particularly interested in in this study. Independent of these considerations, and as discussed before, we use the axisymmetric modeling code {\sc dysmal} for consistency with the studies by \cite{Genzel17, Genzel20}.

The unique advantage of modeling mock observations is that we can compare the modeling results to intrinsic properties of the simulated galaxies. However, in comparing the output of our dynamical modeling to these intrinsic values, it is important to keep in mind some model assumptions. We discuss those assumptions in Section~\ref{s:modelconv} for velocity dispersion, baryonic mass, and central dark matter fraction, before we compare model outputs and intrinsic values in Section~\ref{s:modelcomp}.

\begin{table*}
\caption[]{Comparison for the seven selected TNG50 galaxies seen in five projections each of selected intrinsic properties and dynamical modeling results for our setup assuming modified NFW dark matter halo profiles: for each galaxy we show for $f_{\rm DM}^m(<R_e)=M_{\rm DM}(<R_e)/M_{\rm tot}(<R_e)$, log($M_{\rm bar,20}$), and $\sigma_0$ intrinsic (median) value, mean difference between model output and intrinsic (median) value, and the percentage of lines of sight for which the intrinsic (median) value is recovered within one standard deviation of the marginalized posterior distribution of our MCMC chains. For the intrinsic dark matter fraction $f_{\rm DM,true}$ we list the range of true values for the different model $R_e$. Note that, here, $f_{\rm DM}^m(<R_e)$ is the dark matter mass fraction within a sphere of radius $R_e$. For the intrinsic baryonic mass, we list the sum of stellar and gas mass within 20~kpc. For the velocity dispersion $\sigma_{0,\rm true}$ we list the median of the range of azimuthally averaged `local' values at distances $r=1.5-20$~kpc (see Section~\ref{s:intkin} and Figure~\ref{f:intrvel}). $\Delta\sigma_0$ is the dynamical model $\sigma_0$ minus $\sigma_{0,\rm true}$. In addition, we list the median differences and success rates for all galaxies and lines of sight.}
\label{tab:results}
\centering
\begin{tabular}{lccccccccc}
\hline
    & & $f_{\rm DM}^m(<R_e)$ & && log($M_{\rm bar,20}/M_{\odot}$) && & $\sigma_0$ [km s$^{-1}$] & \\
	ID & $f_{\rm DM,true}^m$ & $\Delta f_{\rm DM}$ & w/in $1\sigma$ & log($M_{\rm bar, true})$ & $\Delta$log($M_{\rm bar})$ & w/in $1\sigma$ & $\sigma_{0,\rm true}$  & $\Delta\sigma_0$  & w/in $1\sigma$ \\
\hline
    \#1 & $[0.24-0.33]$ & $-0.05$ & $100\%$	& 11.19 & $-0.05$	& $80\%$  & 38 & $+18$ & $40\%$ \\
	\#2 & $[0.36-0.40]$ & $-0.10$ &$80\%$   & 11.27 & $-0.01$	& $100\%$ & 26 & $0$ & $60\%$ \\
	\#3 & $[0.34-0.41]$ & $-0.04$ &$60\%$	& 11.40 & $-0.12$ & $40\%$ & 34 & $0$ & $80\%$ \\
	\#4	& $[0.63-0.67]$ & $-0.05$ &$60\%$	& 11.00 & $-0.05$	& $80\%$ & 23 & $+9$ & $80\%$ \\
	\#5	& $[0.33-0.49]$ & $+0.02$ &$20\%$	& 11.24 & $-0.04$	& $40\%$ & 36 & $-2$ & $40\%$ \\
	\#6	& $[0.26-0.43]$ & $+0.03$ &$80\%$	& 10.92 & $-0.12$	& $60\%$ & 21 & $+32$ & $0\%$ \\
	\#7	& $[0.39-0.45]$ & $-0.08$ &$40\%$	& 11.00 & $+0.02$	& $80\%$ & 26 & $+9$ & $20\%$ \\
	\hline
	av && $-0.05$ &$63\%$ && $-0.05$ & $69\%$  && $+3$ & $46\%$ \\
\end{tabular}
\end{table*}

\begin{table*}
\caption[]{Same as in Table~\ref{tab:results} but for a dynamical modeling that assumes pure NFW haloes.}
\label{tab:results_nfw}
\centering
\begin{tabular}{lccccccccc}
\hline
    & & $f_{\rm DM}^m(<R_e)$ & && log($M_{\rm bar,20}/M_{\odot}$) && & $\sigma_0$ [km s$^{-1}$] & \\
	ID & $f_{\rm DM,true}^m$ & $\Delta f_{\rm DM}$ & w/in $1\sigma$ & log($M_{\rm bar, true})$ & $\Delta$log($M_{\rm bar})$ & w/in $1\sigma$ & $\sigma_{0,\rm true}$  & $\Delta\sigma_0$  & w/in $1\sigma$ \\
\hline
    \#1 & $[0.23-0.32]$ & $-0.17$ & $0\%$	& 11.19 & $-0.01$	& $60\%$  & 38 & $+17$ & $40\%$ \\
	\#2 & $[0.35-0.40]$ & $-0.24$ &$0\%$    & 11.27 & $+0.05$	& $100\%$ & 26 & $0$ & $60\%$ \\
	\#3 & $[0.34-0.41]$ & $-0.13$ &$20\%$	& 11.40 & $-0.09$ & $40\%$ & 34 & $0$ & $80\%$ \\
	\#4	& $[0.63-0.67]$ & $-0.08$ &$40\%$	& 11.00 & $-0.02$	& $80\%$ & 23 & $+9$ & $80\%$ \\
	\#5	& $[0.33-0.49]$ & $+0.00$ &$20\%$	& 11.24 & $-0.03$	& $40\%$ & 36 & $-2$ & $20\%$ \\
	\#6	& $[0.26-0.42]$ & $+0.10$ &$60\%$	& 10.92 & $-0.22$	& $40\%$ & 21 & $+32$ & $0\%$ \\
	\#7	& $[0.39-0.44]$ & $-0.09$ &$40\%$	& 11.00 & $-0.01$	& $60\%$ & 26 & $+9$ & $20\%$ \\
	\hline
	av && $-0.14$ &$26\%$ && $-0.03$ & $59\%$  && $+4$ & $43\%$ \\
\end{tabular}
\end{table*}

\subsubsection{Relation between model output and simulation intrinsic measurements}\label{s:modelconv}

(i) For baryonic masses, our model assumes a specific mass distribution of a thick exponential disc and a central bulge, while for our intrinsic measurement we simply sum the baryonic mass of the central galaxy within a sphere of radius $r=20$~kpc. 
For comparing intrinsic and model baryonic masses, we therefore use the model baryonic mass within 20~kpc, $M_{\rm bar,20}$, instead of the total mass integrated to infinity. Typically, 99 per cent of $M_{\rm bar}$ are encompassed in $M_{\rm bar,20}$ (89 per cent for the largest galaxy \#4).
Intrinsically, there might be some amount of additional, extra-planar, stellar or gaseous material within $r=20$~kpc that is not reflected in our dynamical model that assumes a specific mass distribution. Therefore, we would expect model baryonic masses $M_{\rm bar,20}$ that tend to be lower compared to the intrinsic measurement.

(ii) For velocity dispersions, our model assumes an isotropic and constant value throughout the galactic disc, while our intrinsic measurement captures the local velocity dispersion and its azimuthally averaged variations (see Section~\ref{s:intkin}). For a more meaningful comparison, we therefore use the median of the azimuthally averaged, local velocity dispersion at distances $r=1.5-20$~kpc from the center (cf.\, blue lines in Figure~\ref{f:intrvel}) to compare to our model output. Furthermore, our kinematic model neglects any specific motions besides velocity dispersion and in-plane disc rotation, such as inflows, outflows, or warps. Since substantial vertical and radial motions are present in the simulated galaxies (see Section~\ref{s:intkin}), we would expect model velocity dispersions that tend to be higher compared to our intrinsic measurements.

(iii) For dark matter fractions, our primary model output is $f_{\rm DM}^v(<r)=v_{\rm DM}^2(r)/v_{\rm circ}^2(r)$, following \cite{Genzel17, Genzel20}, while our intrinsic measurement gives the fraction of dark matter mass to total mass within a sphere of a certain radius, $f_{\rm DM}^m(<r)=M_{\rm DM}(<r)/M_{\rm tot}(<r)$. 
For a spherical mass distribution, $f_{\rm DM}^v$ and $f_{\rm DM}^m$ are identical. However, in our dynamical models most of the baryonic mass is distributed in a flattened $n_{S}=1$ disc. Therefore, the velocity-based fraction at the baryonic disc effective radius $R_e$, $f_{\rm DM}^v(<R_e)$, is typically lower compared to the mass fraction $f_{\rm DM}^m(<R_e)$ that is agnostic to the inwards distribution of mass -- however, $f_{\rm DM}^v(<R_e)$ can also be larger than $f_{\rm DM}^m(<R_e)$, for instance in case of large bulge fractions, or low $n_{\rm S,disk}$.
For comparing intrinsic and model dark matter fractions, we therefore convert our velocity-based, model dark matter fractions, $f_{\rm DM}^v$, to mass-based dark matter fractions, $f_{\rm DM}^m$, using our best-fit three-dimensional baryonic and dark matter model mass distribution (see also S.~Price, in prep.). On average, $f_{\rm DM}^m(<R_e)$ is larger by a factor of 1.11 (1.09) compared to $f_{\rm DM}^v(<R_e)$ for our TNG50 models with modified (unmodified) NFW haloes.

\subsubsection{Comparison between model output and intrinsic measurements}\label{s:modelcomp}

\begin{figure}
	\centering
	\includegraphics[width=0.95\columnwidth]{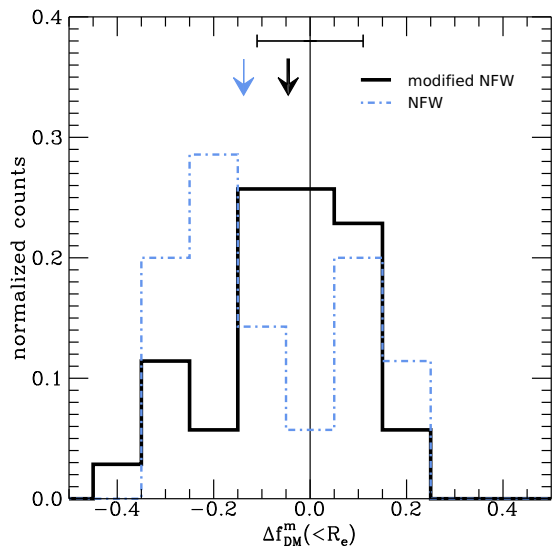}
	\caption[]{Difference in intrinsic {\it vs.\ }inferred central dark matter fractions $f_{\rm DM}^m(<R_e)=M_{\rm DM}(<R_e)/M_{\rm tot}(<R_e)$ based on the dynamical modeling of seven TNG50 simulated galaxies seen from five different projections each. The histograms show the distribution of $\Delta f_{\rm DM}^m(<R_e)= f_{\rm DM,fit}^m(<R_e)-f_{\rm DM,true}^m(<R_e)$ for fits with a modified NFW halo (black) and a pure NFW halo (blue dash-dotted), and the arrows indicate the median $\Delta f_{\rm DM}(<R_e)$. The error bar at the top gives the typical uncertainty on the central dark matter fraction for fits to observational data, $\delta f_{\rm DM}(<R_e)\approx\pm0.12$. For our TNG50 galaxies, the model setup with a modified (i.e.\ contracted) NFW halo slightly underestimates the central dark matter fraction by 5 per cent on average, and the setup with an NFW halo by 14 per cent.}
	\label{f:delta_fdm}
\end{figure}

Considering first the modeling results using a modified NFW halo, we find that those galaxies with the most symmetric rotation curves following the $\Delta\chi^2_{\rm red}$ statistics described in Section~\ref{s:asym} (namely \#6 and \#1; cf.\  Figure~\ref{f:1dloscomp}), also have the most accurate modeling results in terms of recovering the central dark matter fraction $f_{\rm DM}^m(<R_e)=M_{\rm DM}(<R_e)/M_{\rm tot}(<R_e)$ (see Table~\ref{tab:results}). For these galaxies, all lines of sight lead to best-fit values of $f_{\rm DM}^m$ that agree within their uncertainties with the intrinsic values. Similarly, the best-fit models for the galaxy with the least symmetric rotation curve (\#5, cf.\ Figure~\ref{f:1dloscomp}) do worst in estimating the central dark matter fraction, and only one line of sight has a model $f_{\rm DM}^m(<R_e)$ that agrees within its uncertainties with the intrinsic value. 
Overall, we recover $f_{\rm DM}^m(<R_e)$ in $63$ per cent of cases within one standard deviation ($68^{\rm th}$ percentile) of the one-dimensional marginalized MCMC posterior distributions. Note that this success rate is expected and therefore suggests that our derived MCMC uncertainties are realistic estimates.
In the median (mean), the modeling results in central dark matter fractions are lower by $5$ ($4$) per cent compared to $f_{\rm DM,true}^m$. This is well within typical uncertainties on $f_{\rm DM}(<R_e)$ derived from modeling of observational data at high$-z$, which are $10-15$ per cent (see also Figure~\ref{f:delta_fdm}). 

We recover the baryonic mass within 20~kpc in $69$ per cent of cases within one standard deviation of the one-dimensional marginalized MCMC posterior distributions. In the median, $M_{\rm bar,20}$ is slightly underestimated through the modeling (by $0.05$~dex), as expected. The typical observational uncertainty on $z\sim1-2$ baryonic galaxy mass estimates is $0.2$~dex.

We recover the median galaxy-wide ($r=1.5-20$~kpc) intrinsic velocity dispersion in 46 per cent of cases within the uncertainties. In the median, $\sigma_0$ is slightly overestimated through the modeling (by $3$~km~s$^{-1}$), as expected.
We list median differences between intrinsic $f_{\rm DM}^m(<R_e)$, baryonic mass, and $\sigma_0$ in Table~\ref{tab:results} for our modeling setup with modified NFW haloes.

If we model our kinematic extractions with a pure NFW halo instead of the modified, intrinsically constrained $\alpha\neq1$ halo profile, we get slightly worse results regarding the galactic parameters:
the central dark matter fraction is recovered only in $26$ per cent of cases, and the baryonic mass within 20~kpc in 59 per cent of cases. Here, $f_{\rm DM}^m(<R_e)$ is typically underestimated by 14 per cent. We list the corresponding median differences between intrinsic $f_{\rm DM}^m(<R_e)$, baryonic mass, and $\sigma_0$ for fits assuming an NFW halo in Table~\ref{tab:results_nfw}.
We note that the NFW fits taken at face value are not {\it per se} worse. On the contrary, they have comparable $\chi^2$ statistics with a small mean difference of $\chi^2_{\rm red,mNFW}-\chi^2_{\rm red,NFW}=0.29$. This indicates that for the case of our TNG50 galaxies it is not possible to differentiate between the two halo models based on the goodness of fit to the extracted one-dimensional kinematics alone \citep[see also e.g.][]{Pineda17}.

Figure~\ref{f:delta_fdm} illustrates the deviations from the true dark matter fraction for our model setups with a standard and a modified NFW halo. The histograms show the distribution of $\Delta f_{\rm DM}^m= f_{\rm DM,fit}^m-f_{\rm DM,true}^m$ for fits with a modified NFW halo (black) and a pure NFW halo (blue dash-dotted), and the arrows indicate the median $\Delta f_{\rm DM}^m$. 
As in Tables~\ref{tab:results} and \ref{tab:results_nfw}, we consider the dark matter mass fraction within $r=R_e$, where $R_e$ is the model output best-fit baryonic disc effective radius.
While both model setups (slightly) underestimate $f_{\rm DM}^m(<R_e)$ on average, the modified NFW setup performs somewhat better. For the TNG50 galaxies, this is not unexpected: recall from Figure~\ref{f:dmdensity} that the NFW fit generally underestimates the dark matter density within the inner $\sim10$~kpc for those galaxies.\footnote{
This should be even more pronounced when standard concentration parameters were assumed instead of the typically higher values determined from an NFW fit to the simulated data.} 
Assumptions on the halo profile can potentially have a systematic effect on galaxy-scale dynamical masses and total dark matter halo masses derived from observed kinematics.  
\cite{Genzel20} show that the low central dark matter fractions of their massive, high$-z$ SFGs (our comparison sample), can be explained if the associated haloes have central cores (see also S.~Price et al., in prep.) -- this is in contrast to the simulated TNG50 haloes which have steeper inner profiles than standard NFW.

\subsection{Comparison to observations}\label{s:obscomp}

We now compare intrinsic, as well as mock-observed and subsequently modelled properties of the TNG50 sample to the selected G20 galaxies. For this, we average our modeling results using a modified (i.e., contracted) and a standard NFW halo, similar to the observational comparison values by \cite{Genzel20} which are averages of adiabatically contracted and standard NFW haloes.

\subsubsection{Intrinsic velocity dispersion}\label{s:veldisp}

We start with the intrinsic velocity dispersion $\sigma_0$. As discussed in Section~\ref{s:obsinterp}, this is an important quantity in the observational interpretation of the results by \cite{Genzel17, Genzel20} because of its effect on the outer rotation curve shape through possible reduction of rotational speeds via pressure support. In Section~\ref{s:intkin} we showed for the TNG50 sample that different measures of the intrinsic velocity are approximately constant as a function of radius, and that our modeling procedure can recover the intrinsic velocity dispersion within one standard deviation of the marginalized MCMC posterior distribution in most cases (Section~\ref{s:recovery}).

\begin{figure}
	\centering
	\includegraphics[width=\columnwidth]{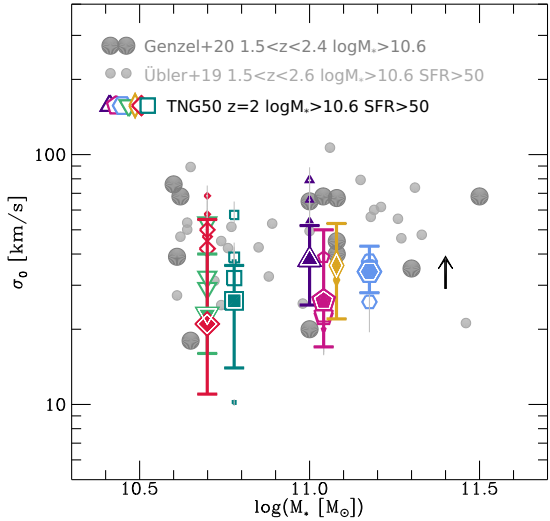}
	\caption{Gas velocity dispersion $\sigma_0$ as a function of intrinsic stellar mass $M_*$ for the simulated galaxies selected from TNG50 at $z=2$ (colored symbols), compared to the observational data by \citet{Genzel20} (large grey circles) and \citet{Uebler19} (small light grey circles; see main text). Constraints from mock-observed and modelled galaxies are shown as open symbols (five per galaxy corresponding to the five lines of sight analysed), where larger sizes indicate better goodness-of-fit. Filled symbols indicate the median of the intrinsic, azimuthally averaged `local' velocity dispersion, and the corresponding error bar indicates the full range of values at distances $r=1.5-20$~kpc. The black arrow approximately indicates (note the log scale) how far the intrinsic median values would increase if a `thermal term' were included in the measurement of the velocity dispersion \citep[see][]{Pillepich19}. Overall, both the intrinsic and mock-observed plus modelled velocity dispersions broadly agree with observations, with on average somewhat lower values.}
	\label{f:veldisp}
\end{figure}

In Figure~\ref{f:veldisp} we show $\sigma_0$ as a function of stellar mass for the intrinsic and modeled simulated data, together with observations from \cite{Genzel20} and \cite{Uebler19}. Intrinsically, the TNG50 sample spans azimuthally-averaged values of $\sigma_0\sim10-55$~km~s$^{-1}$ (colored error bars), with galaxy-wide (excluding the inner $1.5$~kpc) medians ranging between $\sigma_0\sim20-40$~km~s$^{-1}$ (filled symbols; see also Figure~\ref{f:intrvel}). Through modeling of our mock-observations, we recover the galaxy wide median in $>40$ per cent of cases within one standard deviation of the marginalized posterior distribution (open symbols; see Tables~\ref{tab:results} and \ref{tab:results_nfw}). For our TNG50 kinematic sample, there is a tendency to recover slightly larger $\sigma_0$ values from the modeling of the mock-observed galaxies, with velocity dispersions that are on average 9~km~s$^{-1}$ larger than the intrinsic medians, giving an average value of modeled $\sigma_0\sim39$~km~s$^{-1}$. As discussed above, this could be due to the vertical and radial velocity components in the simulated galaxies that are not accounted for by our dynamical model.

The simulation results are compared to the model output by \cite{Genzel20} (large grey circles). 
To give a better sense of typical dispersion values of massive, star-forming discs at this cosmic epoch, we show a subset of $z\geq1.5$ KMOS$^{\rm 3D}$ data by \cite{Uebler19} (small light grey circles). This subset has been selected with the same stellar mass and SFR cuts as the TNG data, and shows a spread of $\sigma_0\sim20-100$~km~s$^{-1}$, with a median of 49~km~s$^{-1}$. Compared to this sample representative of main sequence star-forming discs, the selected TNG50 galaxies lie in the lower half of the observed scatter \citep[see also][]{Vincenzo19}.

For the comparison to the intrinsic values of $\sigma_0$ in the simulations the measurement procedure plays an important role. As described in Section~\ref{s:intkin}, our $\sigma_{0, \rm true}$ ranges give luminosity-weighted azimuthal averages of the `local' velocity dispersion, which is measured in $xy$ bins of 0.5~kpc length. The average of the medians of these measurements at distances $r=1.5-20$~kpc is 29~km~s$^{-1}$. For this way of measuring velocity dispersion, this value is typical of massive log$(M_*/M_{\odot})=10.5-11$ SFGs in TNG50 \citep[see Figure A1, black line, by][]{Pillepich19}.
\cite{Pillepich19} consider including the effects of thermal broadening for the gas velocity dispersion measurement. Their Figures 12 and A1 show that the typical effect for the galaxies studied in this work should be at most of order $+10$~km~s$^{-1}$, which we indicate in our Figure~\ref{f:veldisp} by the black arrow (note the log scale). Including this effect would bring the $\sigma_0$ values (both intrinsic and mock-observed) of the selected TNG50 galaxies in better agreement with the average velocity dispersion of observed SFGs.
In fact, if we apply a Kolmogorov-Smirnov statistic to the subsample by \cite{Uebler19} and the TNG50 dynamical modeling output with the addition of $+10$~km~s$^{-1}$ for thermal broadening, we find that the samples are consistent with being drawn from a common parent sample, whereas without the thermal term they differ by more than $3\sigma$.

\subsubsection{Central dark matter fraction}

\begin{figure*}
	\centering
	\includegraphics[width=0.48\textwidth]{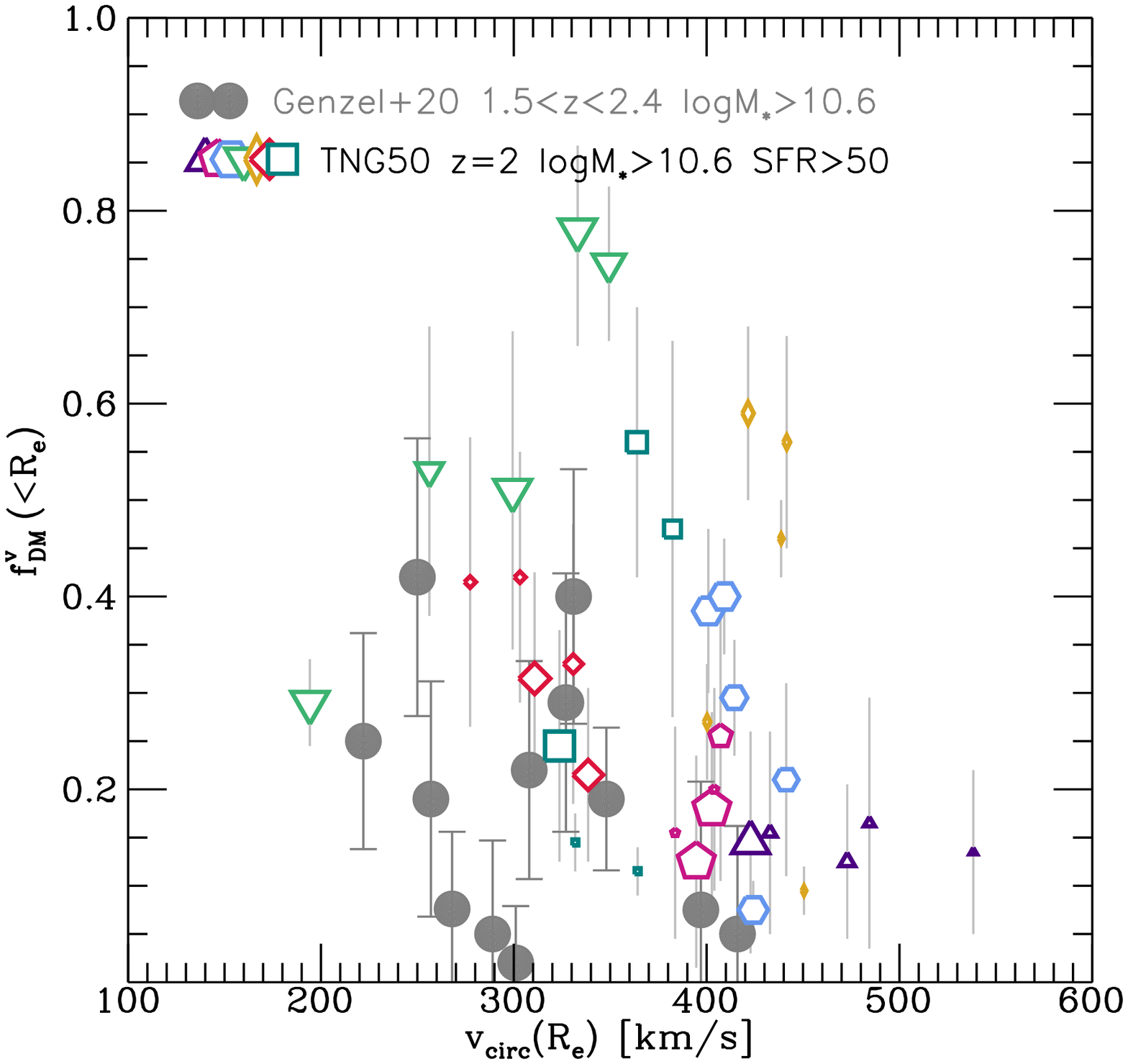}
	\includegraphics[width=0.48\textwidth]{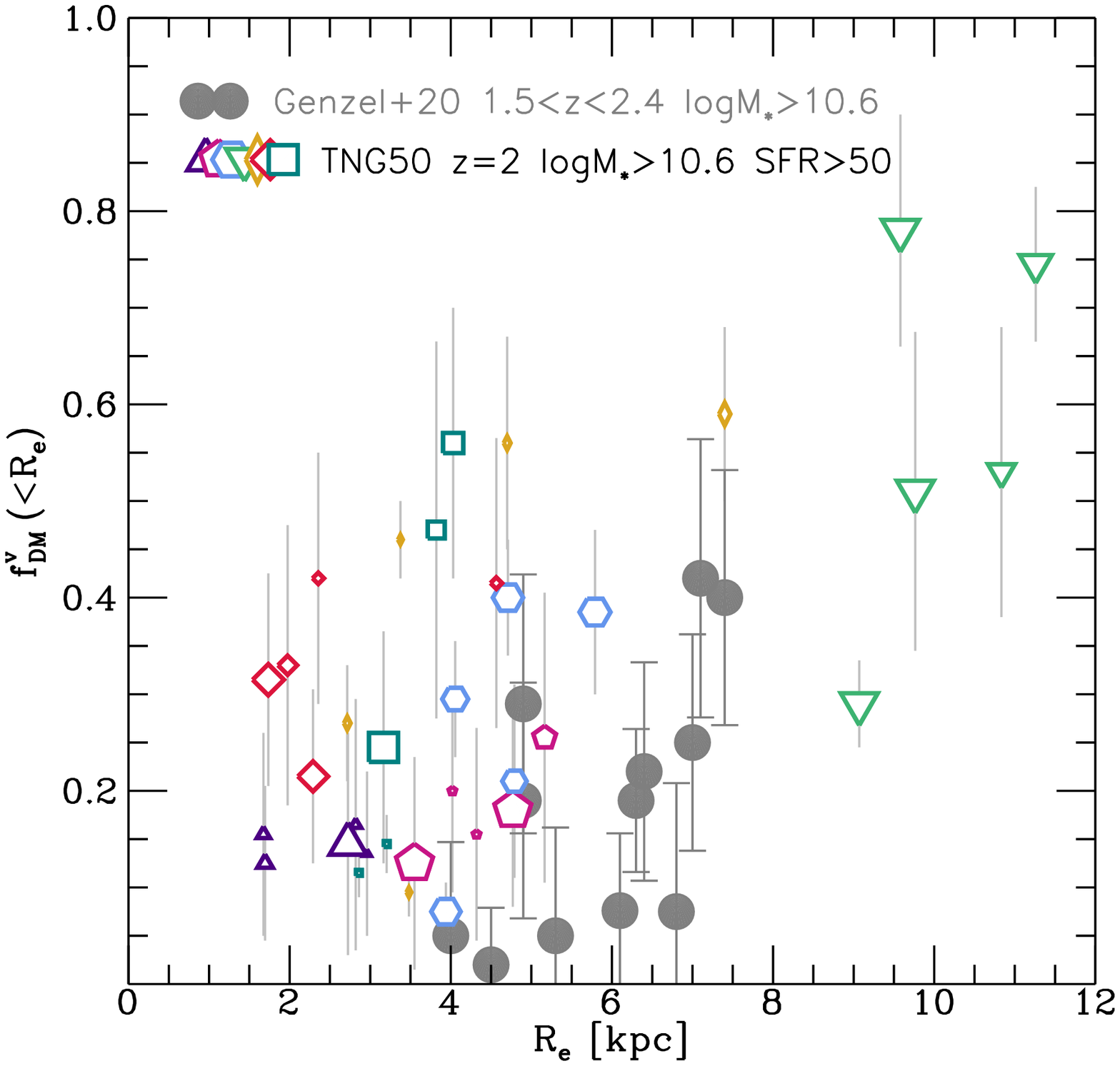}
	\caption{Left: Dark matter fraction within the baryonic disc effective radius $f_{\rm DM}^v(<R_e)=v_{\rm DM}^2(R_e)/v_{\rm circ}^2(R_e)$ as a function of circular velocity at the baryonic disc effective radius $v_{\rm circ}(R_e)$ for the selected simulated TNG50 galaxies at $z=2$ with $M_*>4\times10^{10}M_{\odot}$ and SFR $> 50 M_{\odot}$~yr$^{-1}$ (colored symbols), in comparison to the selected observational data by \citet{Genzel20} (grey circles). For the mock-observed and modelled TNG50 galaxies, larger sizes indicate higher goodness-of-fit (averaged over our setups with modified and unmodified NFW haloes). Within the modeling uncertainties, the simulated and observed populations partly overlap, but the simulated galaxies are offset towards higher velocities and dark matter fractions.
	Right: $f_{\rm DM}^v(<R_e)$ as a function of baryonic disc effective radius $R_e$, with symbols as in the left panel. Here we see a more distinct offset of the simulated and observed galaxies, where at fixed $R_e$ the observed galaxies have smaller dark matter fractions.}
	\label{f:fdm_vcirc}
\end{figure*}

As a final step, we now turn to the dynamical contribution of dark matter on galactic scales.
In the left panel of Figure~\ref{f:fdm_vcirc} we show the velocity-based dark matter fraction within the baryonic disc effective radius $f_{\rm DM}^v(<R_e)=v_{\rm DM}^2(R_e)/v_{\rm circ}^2(R_e)$ as a function of circular velocity $v_{\rm circ}(R_e)$. Grey circles indicate the observations by \cite{Genzel20}, whereas open colored symbols correspond to our dynamical modeling results for the mock-observed TNG50 galaxies.

For our simulated sample, we find values of $f_{\rm DM}^v(<R_e)=0.08-0.78$, and typical uncertainties of $\delta f_{\rm DM}^v(<R_e)\sim0.10$.
34 per cent (12/35) of best-fit dynamical models indicate dark matter fractions of $f_{\rm DM}^v(<R_e)\leq0.2$, and 80 per cent (28/35) $f_{\rm DM}^v(<R_e)\leq0.5$, comparable to the results for the selected G20 sample. 
The average central dark matter fraction of our modeled TNG50 mock observations is $f_{\rm DM}^v(<R_e)\sim0.32$, while for the selected G20 galaxies it is lower by about 40 per cent, with $f_{\rm DM}^v(<R_e)\sim0.19$. However, considering the uncertainties from the dynamical modeling of both real and simulated galaxies, we conclude that the majority of mock-observations are, at face value, 
in broad agreement with the dark matter fractions found by \cite{Genzel20}. In particular, this applies to galaxy \#1, for which all lines of sight give $f_{\rm DM}^v(<R_e)<0.2$. This galaxy has also intrinsically the lowest dark matter fraction.

There is however another important aspect to this comparison. As pointed out in Section~\ref{s:selection}, the simulated galaxies have on average smaller sizes compared to the G20 sample. This is also reflected in the modelling results, as can be seen in the right panel of Figure~\ref{f:fdm_vcirc}, where we show $f_{\rm DM}^v(<R_e)$ as a function of baryonic disc effective radius $R_e$. 
At fixed galactocentric distance, the difference between the dark matter fractions inferred from observations and simulations is evident. Where the dynamical modeling of the mock-observations indicates $f_{\rm DM}^v(<R_e)<0.2$, the corresponding baryonic disc effective radii are always smaller than 5~kpc. In contrast, the low dark matter fractions of the G20 sample are found over a large range in disc sizes, from $R_e\sim4$~kpc to $R_e\sim7$~kpc. 
Importantly, \cite{Genzel20} also find a strong anti-correlation between $f_{\rm DM}(<R_e)$ and baryonic surface density that is qualitatively also found in IllustrisTNG. This suggests that the discrepancy between real and simulated galaxies reported here is likely underestimated, given the smaller sizes of the TNG50 galaxies at comparable masses.

Galaxy \#4 (green triangles) is considerably larger than the other simulated galaxies, with a correspondingly higher dark matter fraction of $f_{\rm DM}^v(<R_e\sim10{\rm{ kpc}})\sim0.6$. Figure~\ref{f:2dmaps} suggests that our $V-$band size prior for this galaxy could be biased high due to extended emission of assembling material -- likely the aftermath of a recent merger (see Section~\ref{s:intkin}). This interaction is likely also responsible for the particularly irregular kinematics (see Figure~\ref{f:1dloscomp}), leading to dynamical estimates of the central dark matter fraction that differ by up to $\Delta f_{\rm DM}^v\sim0.5$.

\begin{figure}
	\centering
	\includegraphics[width=\columnwidth]{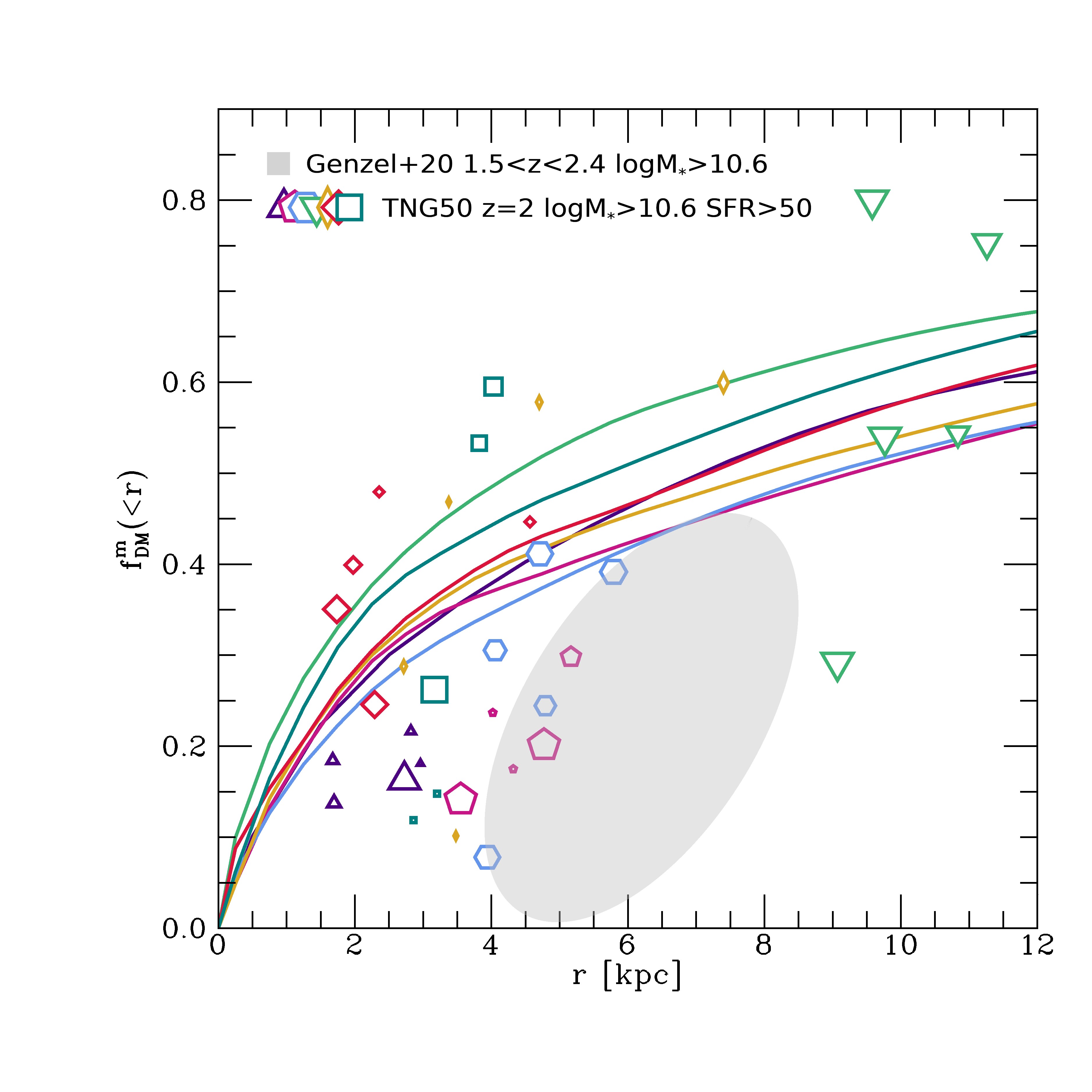}
	\caption{Enclosed dark matter fraction $f_{\rm DM}^m(<r)=M_{\rm DM}(<r)/M_{\rm tot}(<r)$ as a function of radius $r$. Colored lines show the intrinsic profiles of the seven TNG50 galaxies selected for this comparison at $z=2$, and open colored symbols indicate the inferred $f_{\rm DM}^m(<R_e)$ from the modeled mock-observations. The grey shaded area indicates the approximate location of the selected observed galaxies by \citet{Genzel20}. All mock-observed and intrinsic low ($f_{\rm DM}^m(<R_e)<0.2$) dark matter fractions are found for $R_e<4.5$~kpc, in contrast to the observationally inferred values.}
	\label{f:fdm_r}
\end{figure}

We explore the connection between central dark matter fractions and the distances at which they are measured in Figure~\ref{f:fdm_r}. Here, we show as colored lines the intrinsic dark matter fraction as a function of radius for the simulated galaxies. As before, the open colored symbols correspond to the model outputs that are now converted to show the enclosed dark matter mass fractions $f_{\rm DM}^m(<R_e)=M_{\rm DM}(<R_e)/M_{\rm tot}(<R_e)$, as described in Section~\ref{s:modelconv}. By the grey shaded area we indicate the approximate location of the twelve $M_*>4\times10^{10}M_{\odot}$, $z\geq1.5$ SFGs observed by \cite{Genzel20} (which have values of $f_{\rm DM}^m(<R_e)$ that are on average higher by only four per cent, compared to $f_{\rm DM}^v(<R_e)$).
For our TNG50 mock-observations, we find an average value of $f_{\rm DM}^m(<R_e\sim4.5{\rm kpc})\sim0.34\pm0.10$ ($f_{\rm DM}^m(<R_e\sim3.6{\rm kpc})\sim0.27\pm0.10$ without galaxy \#4). For the selected G20 galaxies, instead, we find an average value of $f_{\rm DM}^m(<R_e\sim5.9{\rm kpc})\sim0.20\pm0.10$.
These observations suggest dark matter fractions that are increasing more slowly with radius out to at least the dynamically inferred effective radii (typically $R_e\sim6$~kpc), and all lie below the intrinsic dark matter fraction profiles of the TNG50 sample. Based on our dynamical modeling output for the simulated galaxies, only about half (18/35) of the models would be compatible with a similar profile shape, and, with the exception of one model (for galaxy \#4), would constrain these shallower profiles out to smaller radii. \\

{\bf Comments on sizes: }
For our dynamical modelling we have used the three-dimensional `observed-frame' $H-$band (rest-frame $V-$band) half-light radius as an input prior on the baryonic disc effective radius, mimicking the approach by \cite{Genzel17} (see Section~\ref{s:modeling}). \cite{Genzel20} have either used the same approach as \cite{Genzel17}, or in some cases fixed $R_e\equiv R_{1/2}$.
Obviously, the distance from the center at which a dark matter fraction is measured has an impact on its value. We have explored using different setups with respect to $R_e$ in our dynamical modeling, such as a flat prior with hard bounds of $2-12$~kpc, fixing $R_e$ to the intrinsic baryonic disc half-mass radius based on a bulge-to-disc decomposition of the azimuthally averaged baryonic surface density (average 7.5~kpc), or fixing $R_e$ to half-light sizes measured from random projections of post-processed mock images \citep[average 6.6~kpc; see][for details]{RodriguezGomez19}. 
Consistent with expectations, dynamical models with larger $R_e$ also give larger $f_{\rm DM}(<R_e)$. Considering such model outputs with, for instance,  $R_e\sim6-7.5$~kpc, we find typical values of $f_{\rm DM}^m(<R_e)>0.5$ for the selected TNG50 galaxies. In comparison, the observationally constrained dark matter fractions at these distances are $f_{\rm DM}^m(<R_e)\sim0.1-0.45$ for the selected G20 galaxies. While the lowest $f_{\rm DM}^m(<R_e)$ constrained from the selected TNG50 galaxies are comparable to the highest $f_{\rm DM}^m(<R_e)$ constrained from the selected G20 galaxies at such $R_e$, the average dark matter fractions from dynamical modeling are a factor of about two higher for the selected TNG50 galaxies. 
This is also consistent with the intrinsic dark matter fraction profiles shown in Figure~\ref{f:fdm_r}.\\

\begin{figure}
	\centering
	\includegraphics[width=\columnwidth]{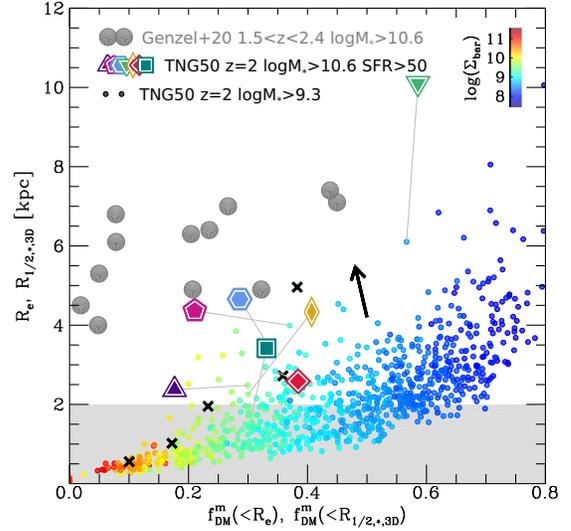}
	\caption{Galaxy size as a function of central dark matter fraction for the observed (grey circles) and simulated (coloured symbols) sample. Large symbols indicate measurements from dynamical modelling, and small circles show intrinsic values. Specifically, for the selected TNG50 galaxies we show model results of $f_{\rm DM}^m(<R_e)$ {\it vs.\ }$R_e$ averaged over five lines of sight and including setups both with and without a modified NFW halo (large coloured symbols). For the full population of $M_*>2\times10^{9}M_{\odot}$ $z=2$ galaxies in TNG50, we show dark matter fractions measured at the three-dimensional stellar half-mass radii $R_{1/2,*,{\rm 3D}}$, colour-coded by their baryonic surface density (small circles).
	The black arrow indicates the average difference between values based on $R_e$ (from our dynamical modeling) {\it vs.\ }$R_{1/2,*,{\rm 3D}}$ (intrinsic to the simulation) for the seven TNG50 galaxies in our kinematic sample (individual galaxies are connected by thin grey lines). Black crosses indicate the location of TNG50 galaxies that meet our SFR and stellar mass cut for modeling, but that were excluded from our kinematic analysis because they are too compact or disturbed. The light grey box roughly indicates the size cut by \citet{Genzel20}. At a fixed central dark matter fraction, the simulated galaxies are typically smaller compared to the selected G20 galaxies.}
	\label{f:r_fdm}
\end{figure}

We further illustrate the offset of mass-based dark matter fractions and sizes in the observed and simulated sample in Figure~\ref{f:r_fdm}. Here we show $f_{\rm DM}^m(<R_e)$ ($f_{\rm DM}^m(<R_{1/2,*,{\rm 3D}})$) on the $x-$axis and $R_e$ ($R_{1/2,*,{\rm 3D}}$) on the $y-$axis. For the modeled TNG50 sample we now plot only one data point per galaxy, which is an average of the fits to the five lines of sight with both a standard and a modified NFW halo (large colored symbols). In addition, we show all $z=2$ TNG50 galaxies (centrals and satellites) with stellar masses $M_* > 2 \times 10^{9} M_{\odot}$ (small colored points). Since we do not have a dynamical measurement of the baryonic disc effective radius for this larger TNG50 sample, we use the three-dimensional stellar half-mass radius $R_{1/2,*,{\rm 3D}}$ and compute the dark matter mass fraction within. Thin grey lines connecting our selected and modeled TNG50 galaxies with the smaller symbols identify the corresponding matches, and the black arrow indicates the average shift for both quantities when going from this simple measurement of dark matter fraction within the $R_{1/2,*,{\rm 3D}}$ to the more complex model output of $f_{\rm DM}^m$ within the dynamically inferred $R_e$. 
At face value, this figures illustrates that at fixed central dark matter fraction, the observed galaxies are larger by factors of $4-14$ on average. In selecting the most massive and highly star-forming systems from the TNG50 $z=2$ snapshot, i.e. those simulated galaxies corresponding most closely in mass and SFR to our G20 reference selection, and by measuring at $R_e$ inferred from dynamical modeling, this stark difference reduces to factors of $\lesssim2$ for dynamically modeled and averaged lines of sight. This underlines the importance of sample selection and analysis techniques.

The color-coding of the $z=2$, $M_*>2\times10^{9}M_{\odot}$ TNG50 population indicates their baryonic surface density within the three-dimensional stellar half-mass radius. \cite{Genzel20} find a steep correlation between central dark matter fraction and baryonic surface density with typical values in the range log$(\Sigma_{\rm bar}/(M_{\odot} {\rm kpc^{-2}}))\approx8-9.5$, which is qualitatively also seen in TNG100 (see their Figure 8). Similar to TNG100, we find that also in TNG50, very low dark matter fractions ($f_{\rm DM}^m(<R_e)<0.2$) are found only for very compact systems with log$(\Sigma_{\rm bar}/(M_{\odot} {\rm kpc^{-2}}))>9.5$.

We remind the reader that the observed galaxies are selected to mostly have larger than average sizes by requiring spatially well resolved systems (see Figure~\ref{fig:selection}). In Figure~\ref{f:r_fdm}, we roughly indicate by the grey box the parameter space that is therefore not probed by observations in the G20 sample \citep[see][for details]{Genzel20}.
Intrinsically, only one the TNG50 galaxies selected by $M_* > 4 \times 10^{10} M_{\odot}$ and SFR $> 50 M_{\odot}$~yr$^{-1}$ lies substantially above the observed $M-R$ relation by \cite{vdWel14b} with a rest-frame $V-$band half-light size of $R_{1/2}=9.4$~kpc (\#4, green triangle). Its intrinsic dark matter fraction $f_{\rm DM}^m(<R_e)$ is about $0.6$. The model-derived, mass-based dark matter fractions for this galaxy range from 0.29 for one line of sight to $0.53-0.80$ for the other four lines of sight, at best-fit $R_e\sim9-11$~kpc. The large range in derived dark matter fractions is due to the galaxy's major axis kinematics that are very line-of-sight dependent (see upper right panel in Figure~\ref{f:1dloscomp}). Intrinsically, at a distance from the center of $\geq6$~kpc all seven simulated galaxies have dark matter fractions of $f_{\rm DM}^m>0.4$.

Ideally we would need a larger sample of high$-z$, high-resolution simulated galaxies that feature extended discs such as the observational sample, to understand if the large but high-$f_{\rm DM}^m$ galaxy \#4 is characteristic for the IllustrisTNG model, or not.
On the other hand, future instrumentation with increased sensitivity will help to determine if there is a larger population of low-surface brightness SFGs with higher dark matter fractions, that might be missing in state-of-the-art observational samples \citep[see][for current trends and parameter spaces]{Price16, Price19, WuytsS16, Genzel20}. However, we would expect such galaxies to likely have lower stellar masses and SFRs than included in our sample selection.

\section{Summary and Discussion}\label{s:conclusions}

We have studied the detailed kinematics of seven massive, $z=2$ SFGs from the TNG50 simulation. Our focus was on the observational perspective and the comparison to a selection of twelve massive, high$-z$ SFGs observed by \cite{Genzel17, Genzel20}. 
These galaxies are particularly interesting from a theoretical point of view because their inferred dark matter content is in tension with current galaxy formation models.
We created mock observations from five projections for each simulated galaxy including effects of the instrumental PSF and LSF, discretization into pixels, and realistic, random as well as systematic noise. We applied standard observational tools for kinematic analysis and dynamical modeling, specifically the same tools used for the analysis of real galaxies by \cite{Genzel20}.

Such accurate comparisons including all relevant observational effects, and possible systematics due to analysis tools, are crucial to highlight real differences between observations and simulations. This lays the foundation to further constrain physical models entering state-of-the-art cosmological simulations.

We also emphasize that our conclusions are limited by the small sample of TNG50 galaxies that meet our selection criteria of massive $z=2$ SFGs, and physical differences that may remain as a consequence. \\

{\bf Global intrinsic properties.} The simulated galaxies lie in a similar parameter space of $M_*$, SFR, $R_{1/2}$, and gas-to-stellar mass ratios compared to the \cite{Genzel20} sample (Section~\ref{s:selection}; Figure~\ref{fig:selection}), with the former two by selection.
Small differences in these global properties of the observed and simulated samples should not be over-interpreted, given the low-number statistics in both the observational and simulated samples, and the known systematic differences in SFRs and sizes between observations and the TNG model \citep{Genel18, Donnari19, Pillepich19}.

The intrinsic dark matter halo profiles of the TNG50 galaxies are steeper than NFW, possibly due to adiabatic contraction, with typical inner slopes of $\alpha\sim1.6$ (Section~\ref{s:dmdensity}; Figure~\ref{f:dmdensity}; see also \citealp{Lovell18}). 

{\bf Intrinsic kinematics and the role of pressure support.} The intrinsic, azimuthally averaged gas rotation velocities of the TNG50 sample are flat on galactic scales. 
Falling intrinsic rotation curves are only seen for three of the seven selected galaxies (\#5, \#6, \#7) at distances $r\gtrsim7-10$~kpc, beyond the visible extent of the galaxies.
The azimuthal averages of the luminosity-weighted, local gas velocity dispersions at $r>1.5$~kpc are fairly constant with radius, with values $<50$~km~s$^{-1}$. 
All galaxies show substantial vertical and radial motions, with values of $|v_r|$ and $|v_z|\sim50-200$~km~s$^{-1}$ (Section~\ref{s:intkin}; Figure~\ref{f:intrvel}).

As a consequence of the somewhat low velocity dispersions, the effects of pressure support based on the ansatz by \cite{Burkert10} are not very important for the TNG50 galaxies. 
The \cite{Burkert10} pressure correction assumes exponential profiles, hydrostatic equilibrium, axisymmetry and an isotropic velocity dispersion which is independent of radius, and consequently further correction factors would have to be applied if any of these assumptions do not hold. While high-$S/N$, high-resolution observations appear consistent with these assumptions (i.e.\ only small asymmetries, an isotropic velocity dispersion, and an exponential profile), at least the assumption of axisymmetry is violated for the majority of the TNG50 galaxies.\footnote{
Particularly for higher-$S/N$ data of local galaxies more complex approaches are sometimes used to correct for pressure support \citep[e.g.][]{Weijmans08, Oh11, Iorio17} which can be applied to simulated data as well \citep[e.g.][]{Oman19}.}

As an example of a different galaxy formation model, the higher-resolution FIRE-2 simulations \citep{Hopkins14, Hopkins18} do find higher velocity dispersions ($\sigma_0\approx100-150$~km~s$^{-1}$) and therefore a more important role of pressure support in simulated disc galaxies at $1.5<z<3$ \citep{Wellons19}. 
Keeping in mind that alternative operational definitions of gas velocity dispersion may imply a factor of $2-3$ differences in values \citep[see][]{Pillepich19}, we speculate the primary difference in gas velocity dispersions (and their kinematic impact) of massive, high$-z$ main sequence SFGs in the IllustrisTNG and FIRE-2 models to be related to the different implementation of feedback: in IllustrisTNG, stellar feedback-driven winds are hydrodynamically decoupled from the interstellar medium until they escape the galaxy, following \cite{Springel03}, whereas in FIRE-2, mechanical feedback from stars couples directly to the surrounding medium \citep{Hopkins18a}. 
On the other hand, the AGN feedback in IllustrisTNG is directly coupled to the gas, with energy injection from the central super massive black holes directly affecting the coldest and densest gas in galaxies, a feedback channel not included in the FIRE-2 model.

{\bf Mock-observed rotation curve shapes.} 
We construct rotation curves for all simulated galaxies. Along individual lines of sight, however, most simulated galaxies display substantial asymmetries in their rotation curve shapes, such that outer kinematics may differ by up to $\Delta v=200$~km~s$^{-1}$ (Figures~\ref{f:1dlos} and \ref{f:1dloscomp}). These asymmetries are likely caused by minor mergers, as indicated by correlated vertical and radial gas motions with respect to the disc plane, or by tidal features through interaction with nearby galaxies (Section~\ref{s:intkin}; Figure~\ref{f:intrvel}).

Quantifying the asymmetries of the TNG50 rotation curves and comparing them to the observations by \cite{Genzel20}, we find that the simulated galaxies have less regular kinematics (Section~\ref{s:asym}). This is likely connected to the large radial and/or vertical motions we intrinsically find for all galaxies.

{\bf Dynamical modeling.} The success of our dynamical models in recovering intrinsic parameters, and in particular the central dark matter fraction $f_{\rm DM}(<R_e)$, depends on two main factors: (i) assumptions on the inner halo profile, and (ii) the regularity of the galaxy kinematics. If we assume a modified (contracted) NFW halo for which we have constrained the inner slope from fits to the intrinsic dark matter density (Section~\ref{s:dmdensity}; Figure~\ref{f:dmdensity}), we recover $f_{\rm DM}^m(<R_e)=M_{\rm DM}(<R_e)/M_{\rm tot}(<R_e)$ in 63 per cent of cases within one standard deviation of the MCMC posterior distribution. Using a standard NFW halo, we are successful only in 26 per cent of cases. 
For our sample of TNG50 galaxies, the choice of modeling with a standard or a more contracted NFW halo results in an average shift in the inferred DM fraction of about $-0.09$ (Figure~\ref{f:delta_fdm}). 
For real galaxies, of course, the dark matter density profile is typically not known, but our results encourage dynamical modeling with variable, or varying, dark matter density profiles. 
Apart from the halo profile, we see some correlation between the reflection symmetry of rotation curves and the ability of our models to accurately recover the central dark matter fraction: for the two galaxies with the most symmetric rotation curves our modified NFW setup correctly recovers the intrinsic $f_{\rm DM}^m(<R_e)$ output by the simulation for all lines of sight, while for the galaxy with the most asymmetric rotation curves we find only one best-fit (of five) recovering the intrinsic value (Section~\ref{s:recovery}).

{\bf Comparison to observations.} For our comparison to the selected observational results by \cite{Genzel20}, we average the model results using a modified and a standard NFW halo. About 34 (86) per cent of the thus model-derived central dark matter fractions of the TNG50 galaxies have values that are similar, namely $f_{\rm DM}^v(<R_e)<0.2$ ($f_{\rm DM}^v(<R_e)<0.5$), compared to the results by \cite{Genzel20}. 
On average, however, the mean central dark matter fraction of the TNG50 galaxies, $f_{\rm DM}^v(<R_e)\sim0.32\pm0.10$, is larger than that of the selected observational sample by \cite{Genzel20} by a factor of 2 (Section~\ref{s:obscomp}; Figure~\ref{f:fdm_vcirc}). 
This result becomes more substantiated when comparing the galactocentric distances at which the dark matter fractions are measured (Figures~\ref{f:fdm_vcirc} and \ref{f:r_fdm}): for the TNG50 galaxies, we typically find dynamically constrained baryonic disc effective radii $R_e<5$~kpc, and particularly all low dark matter fractions ($f_{\rm DM}^v(<R_e)<0.2$) are found at $R_e<5$~kpc.
This is in contrast to the observations by \cite{Genzel20}, where the average value for high$-z$, massive SFGs, $f_{\rm DM}^v(<R_e)\sim0.2$, is typically measured at $R_e\sim6$~kpc. 
Taking into account different definitions of input priors on galactic sizes and their effect, we find that the mass- and SFR-matched $z=2$ TNG50 galaxies are generally too compact and/or too dark matter-dominated (Figures~\ref{f:fdm_vcirc}, \ref{f:fdm_r}, and \ref{f:r_fdm}). 

Similar results have also been found for $z=0$ galaxies. For instance, a recent study by \cite{Marasco20} shows that massive disc galaxies from the EAGLE and TNG100 simulations live in dark matter haloes that are on average factors of four and two more massive than what has been inferred for corresponding galaxies in the SPARC \citep{Lelli16} sample \citep{Posti19}. 
In a comparison of dark matter fractions for different galaxy types and at different redshifts with TNG100 predictions, \cite{Lovell18} find either broad agreement (e.g.\ with the results from $z=0$ disc galaxies compiled by \citealp{Courteau15} or from $z=0$ early-type galaxies by \citealp{Alabi17} within fixed apertures), or too high dark matter fractions in TNG100 when comparing to observational data (e.g.\ $z=0$ early-type galaxies by \citealp{Alabi17} within five times the effective radii and by \citealp{Cappellari13} within one effective radius). These authors also note that some of the latter discrepancies would be alleviated if the simulated haloes would not contract due to the presence of baryons.\\

The massive $z=2$ TNG50 SFGs analysed in this paper differ from real galaxies observed by \cite{Genzel17, Genzel20} specifically in their dark matter fractions at the dynamically constrained baryonic disc effective radius. Quantitatively, at fixed $R_e$ the TNG50 dark matter fractions are too high by a factor of about 2. 

We speculate that this may be due to physical processes which are not resolved in sufficient detail with the numerical resolution available in current cosmological simulations.
At $z\sim2$, during the peak epoch of cosmic star formation rate density, galaxies are subject to rapid baryon assembly, wide-spread condensation of gas into stars \citep[e.g.][]{Whitaker14}, and dissipative processes due to large gas fractions \citep[e.g.][]{Tacconi18}. 
In addition, galaxies are shaped through stellar feedback-driven outflows, increasing with SFR, and the high duty cycle of active galactic nuclei-driven outflows at high masses \citep[e.g.][]{FS19}.
From the theoretical side, there is no final consensus on the implementation of the relevant physical processes via sub-grid recipes \citep[cf.][]{Naab17}. 
This becomes particularly evident when considering the stark variations in kinematics of simulated high$-z$ SFGs realized through different models of galaxy formation \citep[e.g.\ this work;][]{Pillepich18a, Pillepich19, Teklu18, Wellons19}.  

The recent observational results by \cite{Genzel20} highlight for the first time at $z\gtrsim1$ the coupling between central baryonic surface densities and dark matter fractions on galaxy scales. This might point toward efficient heating of the galaxy-scale dark matter halo due to dynamical friction and/or strong feedback -- processes that might not be sufficiently resolved or appropriately modelled by current state-of-the-art cosmological simulations.

The observational findings and the differences between observed and simulated kinematics and dark matter contents carved out in this work encourage future model improvements and comparisons.
From the observational side it would be helpful to have representative measurements of the gas content and distribution for individual galaxies for which H$\alpha$ kinematic observations exist as well. This would allow to investigate in more detail, for instance, if simulated dark matter fractions are too high and galaxies are too compact due to a lack of galactic gas content at high redshift.
Finally, in this paper we have focused on the high-mass end of the star-forming main sequence at $z\sim2$. It would be interesting to expand upon the present work by e.g. including lower-mass or lower-redshift galaxies.

\section*{Acknowledgements}
We are grateful to the anonymous referee for a very constructive report that helped to improve the quality of this manuscript. 
We thank Thorsten Naab, Volker Springel, Rainer Weinberger, and Sarah Wellons for comments on a draft version of this paper. 
H\"U thanks the Center for Computational Astrophysics at the Flatiron Institute for their hospitality. The Flatiron Institute is supported by the Simons Foundation. 
This research was supported by DFG/DIP grant STE/1869 2-1 / GE 625/17-1, and by the Excellence Cluster ORIGINS which is funded by the Deutsche Forschungsgemeinschaft (DFG, German Research Foundation) under Germany's Excellence Strategy -- EXC 2094 -- 390783311. 
The TNG50 simulation was run with compute time granted by the Gauss Centre for Supercomputing (GCS) under GCS Large-Scale Project GCS-DWAR on the Hazel Hen supercomputer at the High Performance Computing Center Stuttgart (HLRS).

\section*{Data availability} 
Data  directly  related  to  this  publication  and  its  figures  is available on reasonable request to the corresponding author. The IllustrisTNG simulations are publicly available and accessible at \url{www.tng-project.org/data} \citep{NelsonD19a}, where the TNG50 simulation will also be made public in the future.

\bibliographystyle{mn2e}
\bibliography{literature}

\begin{appendix}

\section{Dismissed galaxies}
\label{s:dismissed}

\begin{figure*}
	\centering
	\includegraphics[width=0.8\textwidth]{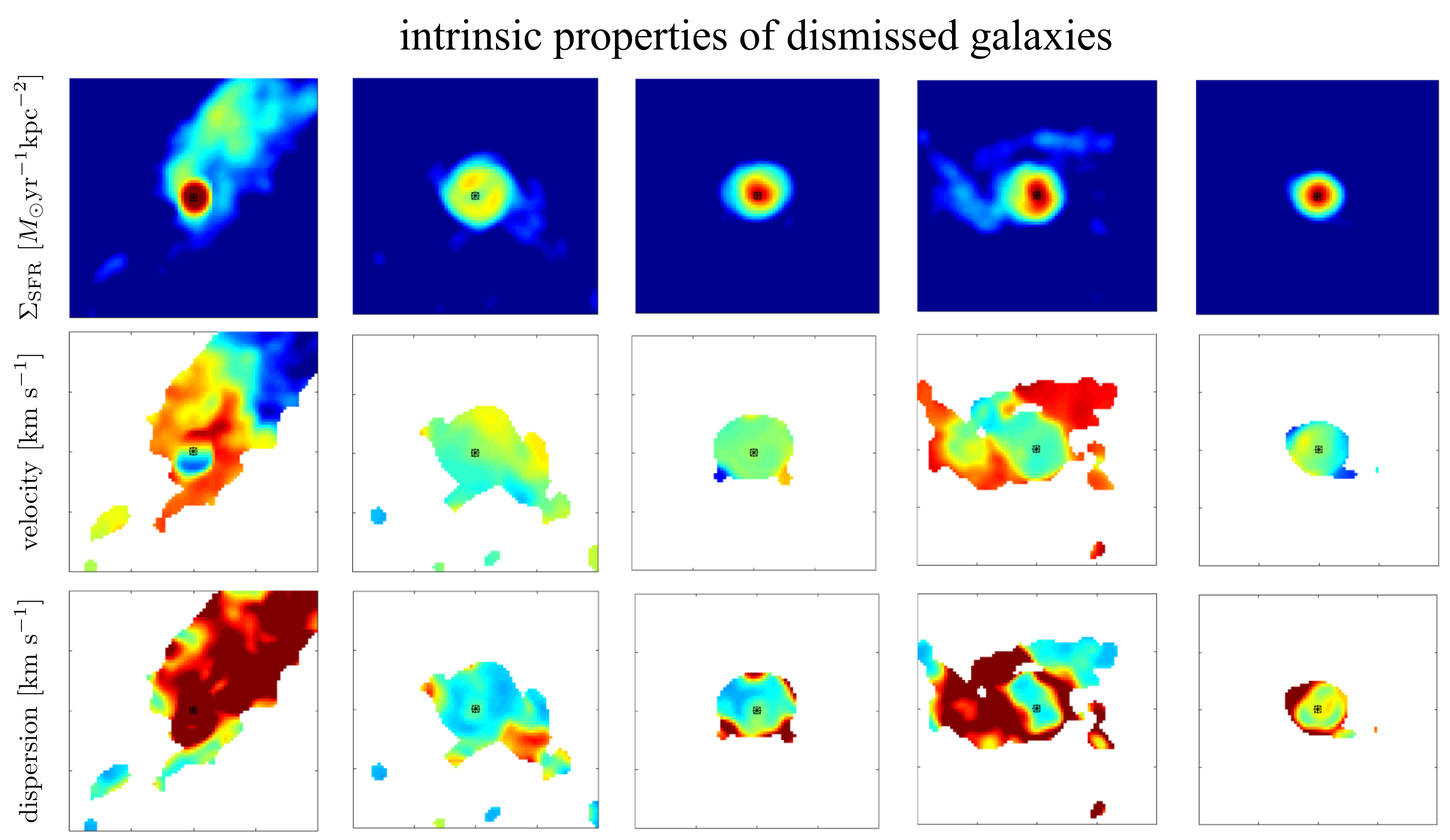}
	\caption[Projected two-dimensional kinematic maps of dismissed galaxies]{Projected two-dimensional maps of the convolved intrinsic parameters (top row: $\Sigma_{\rm SFR}$; second row: velocity; third row: velocity dispersion) for the five dismissed TNG50 galaxies (columns). The projections correspond to face-on, i.e.\ an inclination of $i=0^{\circ}$. The panels show 40~kpc~$\times$~40~kpc in projection, and the color scale shows [-400; 400]~km~s$^{-1}$ for velocity, and [0; 150]~km~s$^{-1}$ for velocity dispersion. The galaxies were dismissed due to strong interaction/disturbance signatures and/or because they were too compact for a kinematic analysis meeting the purpose of this work.}
	\label{f:dismissed}
\end{figure*}

In Figure~\ref{f:dismissed} we show those 5/12 TNG50 galaxies that met the stellar mass and SFR selection cuts (see Table~\ref{tab:tngprop}), but were dismissed from further kinematic analysis because they show signatures of strong interaction or disturbance, and/or they are too compact to extract extended rotation curves (such systems would also be excluded from observational studies for the same reasons).
The galaxies in columns 1-4 all have a similarly massive galaxy (mass ratio $\geq$ 1:2) in their vicinity ($\Delta r=30-60$~kpc), and particularly the first object shows a high-surface brightness accretion or tidal stream.
The galaxy in column 5 is undisturbed, but also the most compact object meeting our other selection criteria. Similarly, the other four objects are very compact in addition to their disturbed kinematics.

\section{Using a more complex PSF model}
\label{ap:2comp}

We explore the effect of using a more complex PSF model for our mock observations on the kinematic extractions and dynamical modeling for galaxy \#3. We repeat the creation of our mock data cubes, the kinematic extraction, and the dynamical modeling as described in Sections~\ref{s:mock}, \ref{s:extractions}, and \ref{s:modeling}, adopting the double Gaussian model fit to the effective, average adaptive optics PSF of SINFONI as measured by \cite{FS18}. 

\begin{figure}
	\centering
	\includegraphics[width=0.95\columnwidth]{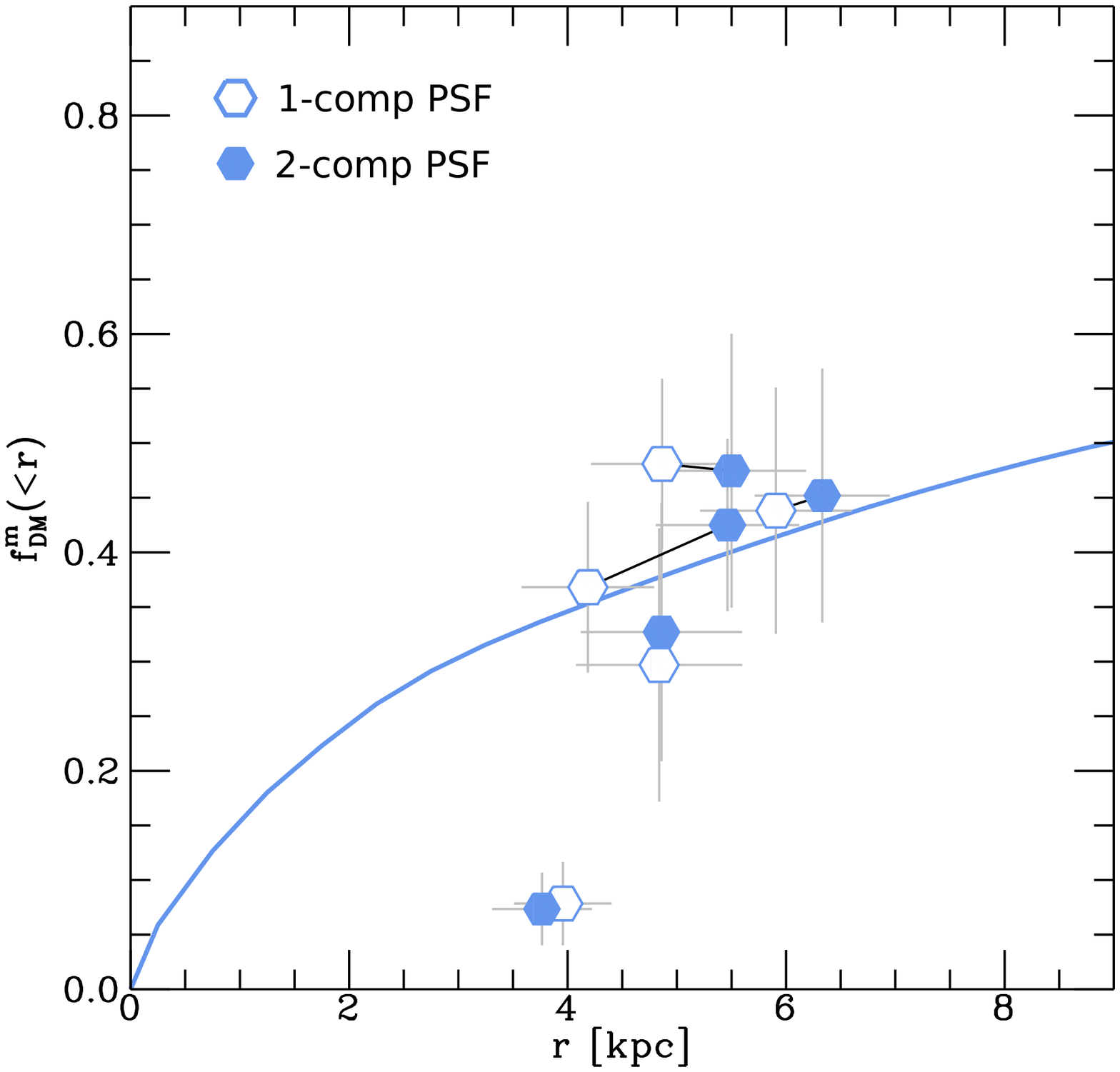}
	\caption{Effect of using a one-component {\it vs.} two-component Gaussian PSF model on our dynamical modeling results for galaxy \#3. The solid line shows the intrinsic enclosed dark matter fraction $f_{\rm DM}^{m}(<r)$ as a function of radius $r$. Open symbols show $f_{\rm DM}^{m}(<R_e)$ {\it vs.} $R_e$ for our model setup using a modified NFW halo and a one-component Gaussian PSF for the five lines of sight. Filled symbols show the corresponding (connected by black lines) results for the two-component Gaussian PSF model as derived by \citet{FS18}. For all lines of sight, the derived central dark matter fractions from both models agree within their uncertainties.}
	\label{f:psf2comp}
\end{figure}

Overall, if we use the same noise scaling as for the original mock cube (for comparison), we are able to trace the emission to somewhat larger distances due to additional smoothing from the seeing-limited halo of the two-component PSF model. Similarly, some noise features are washed out.
Despite these differences, the dynamical modeling results typically agree well within the uncertainties with our original one-component PSF modeling results. In particular, this holds for the central dark matter fraction $f_{\rm DM}(<R_e)$, as is illustrated in Figure~\ref{f:psf2comp}.

\section{Effect of dust on extracted kinematics}
\label{ap:dust}

\begin{figure*}
	\centering
	\includegraphics[width=0.8\textwidth]{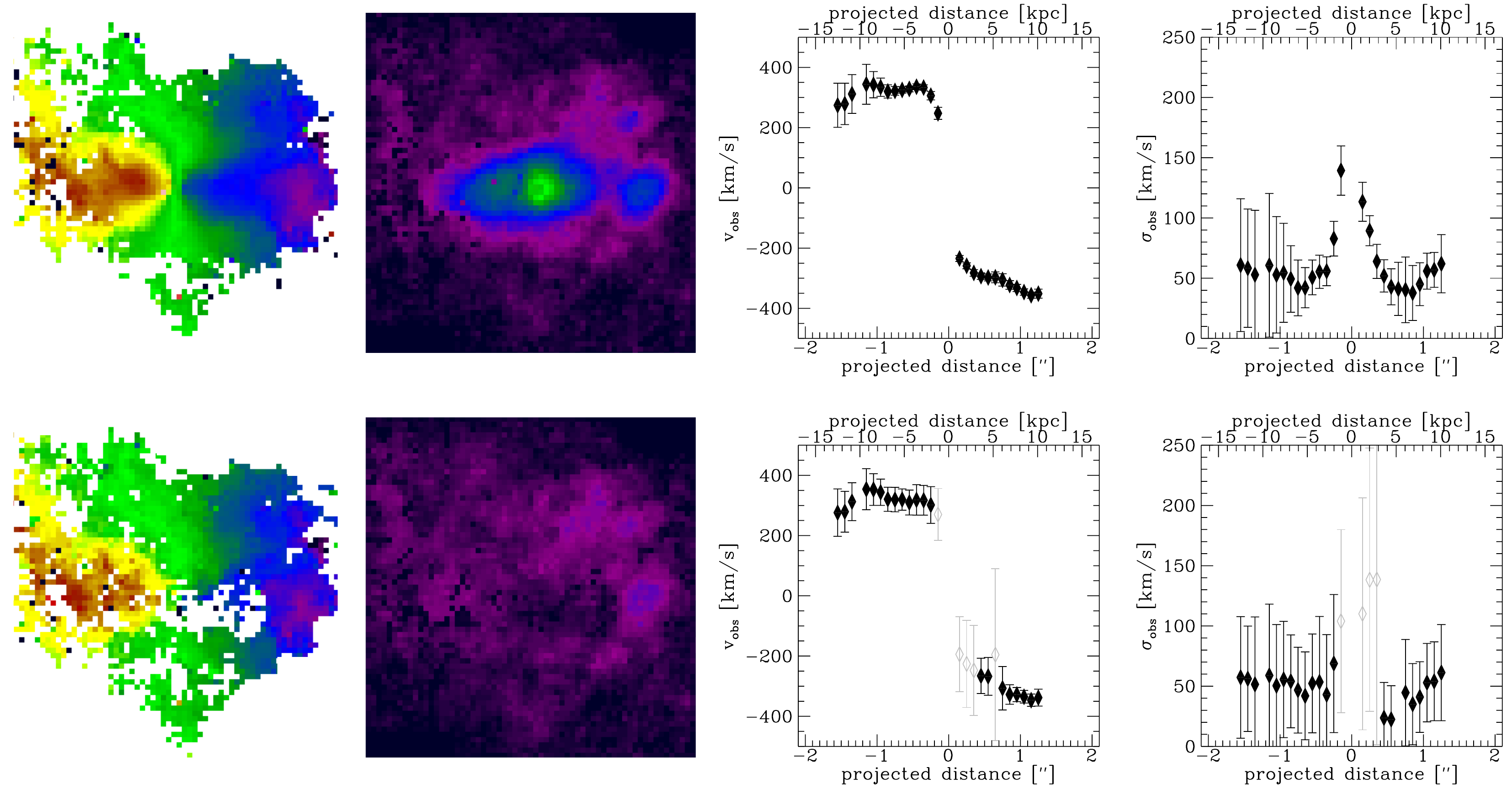}
	\caption[Effects of dust]{Illustrating the effect of dust on the $S/N$ and extracted kinematics for galaxy \#3, sightline 3. From left to right: velocity map showing pixels with $S/N>3$ and color scale [-400; 400]~km~s$^{-1}$; corresponding $S/N$ map; extracted one-dimensional rotation velocity and velocity dispersion for mock-observations without (top) and with (bottom) dust, following the method by \cite{NelsonD18}. See text for details.}
	\label{f:dust}
\end{figure*}

We explore the effect of accounting for dust in our mock-observations and kinematic extractions by using a model that accounts for spatially-resolved dust attenuation. The implementation follows the fiducial dust model (C) by \cite{NelsonD18}\footnote{Neglecting, however, the spatially-unresolved component used in their model (B), which applies for young stars, not for gas.} and makes use of results by \cite{Cardelli89,Calzetti94}. We refer the reader to \cite{NelsonD18} for details.

Since the inclusion of dust extinction decreases the simulated `emission' from the system, we have adjusted the noise level such that sufficiently large $S/N$ is reached up to radii comparable to those presented in the main analysis. This choice enables a direct comparison of the extracted kinematics. In Figure~\ref{f:dust} we show for sightline 3 of galaxy \#3 the mock-observed velocity map, corresponding $S/N$ map, and one-dimensional major axis velocity and dispersion profiles. The bottom row includes dust. In this galaxy as well as in the full sample, particularly the star-forming galaxy centers are affected by dust. Accounting for dust can lead to a more ring-like appearance of the systems, as illustrated in the $S/N$ maps (such structures are also observed in real galaxies; see e.g. \citealp{Genzel08}). With our noise scaling of choice, the overall $S/N$ is much lower in the mock-observation including dust. This is transferred also to the one-dimensional kinematic extractions: there are fewer reliable data points, and the uncertainties are larger. For a direct comparison, we plot in the bottom right panels of Figure~\ref{f:dust} extractions at the same distances from the center as in the top right panels, but we show unreliable extractions in grey. 
We emphasize that the overall kinematic properties of our TNG50 galaxies do not change when accounting for dust, however the $S/N$ and therefore the quality of the kinematic extractions are affected. Particularly the line widths translating into velocity dispersion are more sensitive to this decrease in $S/N$.

\section{Non-circular motions and kinematic asymmetries}
\label{ap:asym}

In Section~\ref{s:intkin}, we discussed the substantial vertical and radial motions in our sample of TNG50 SFGs. Here, we want to briefly demonstrate by reference to one example the effect of artificially removing these components on the extracted kinematics, which we call `equilibration'. 
This procedure exploits the full knowledge about the simulated data and aims at evaluating the effect of non-rotational motions on the regularity of the extracted kinematics. Specifically, the method artificially removes vertical and radial velocity components of the star-forming gas. To achieve this, the galaxy is divided into circular $0.5$~kpc-sized regions, inside each of which the mean radial and vertical velocity is subtracted from each resolution element. This results in no impact on the tangential velocity and a minimal impact on the velocity dispersion. 

\begin{figure}
	\centering
	\includegraphics[width=0.9\columnwidth]{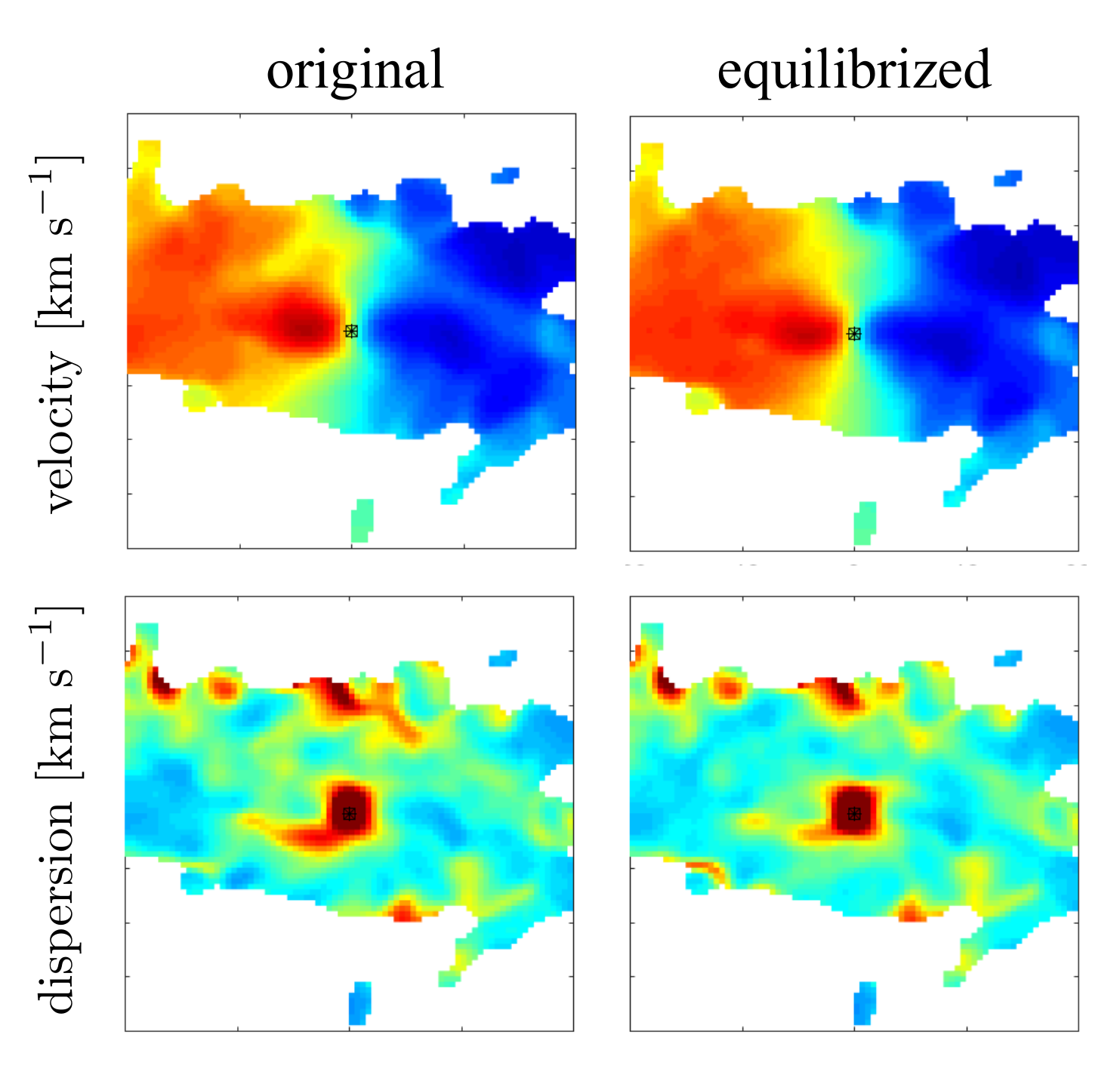}
	\caption{Projected velocity and velocity dispersion maps before (left) and after (right) removing velocity components vertical and radial with respect to the disc plane (`equilibrized') for galaxy \#3, sightline 2. The projections correspond to an inclination of $i=60^{\circ}$. The panels show 40~kpc~$\times$~40~kpc in projection, and the color scale shows [-400; 400]~km~s$^{-1}$ for velocity, and [0; 150]~km~s$^{-1}$ for velocity dispersion. Both the velocity and velocity dispersion fields become smoother and more regular with the vertical motions removed.}
	\label{f:fiddling2d}
\end{figure}

\begin{figure}
	\centering
	\includegraphics[width=0.49\columnwidth]{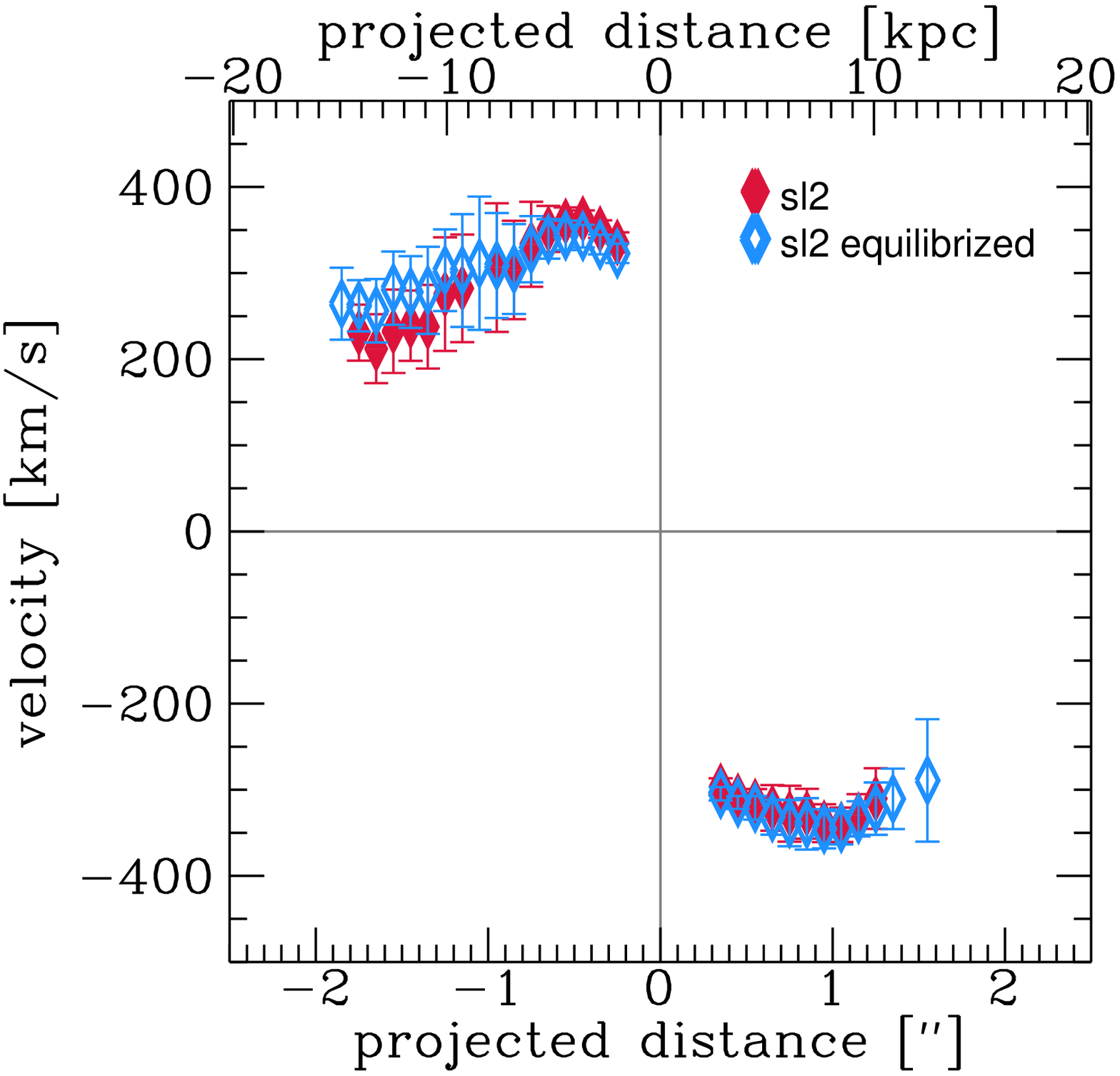}
	\includegraphics[width=0.49\columnwidth]{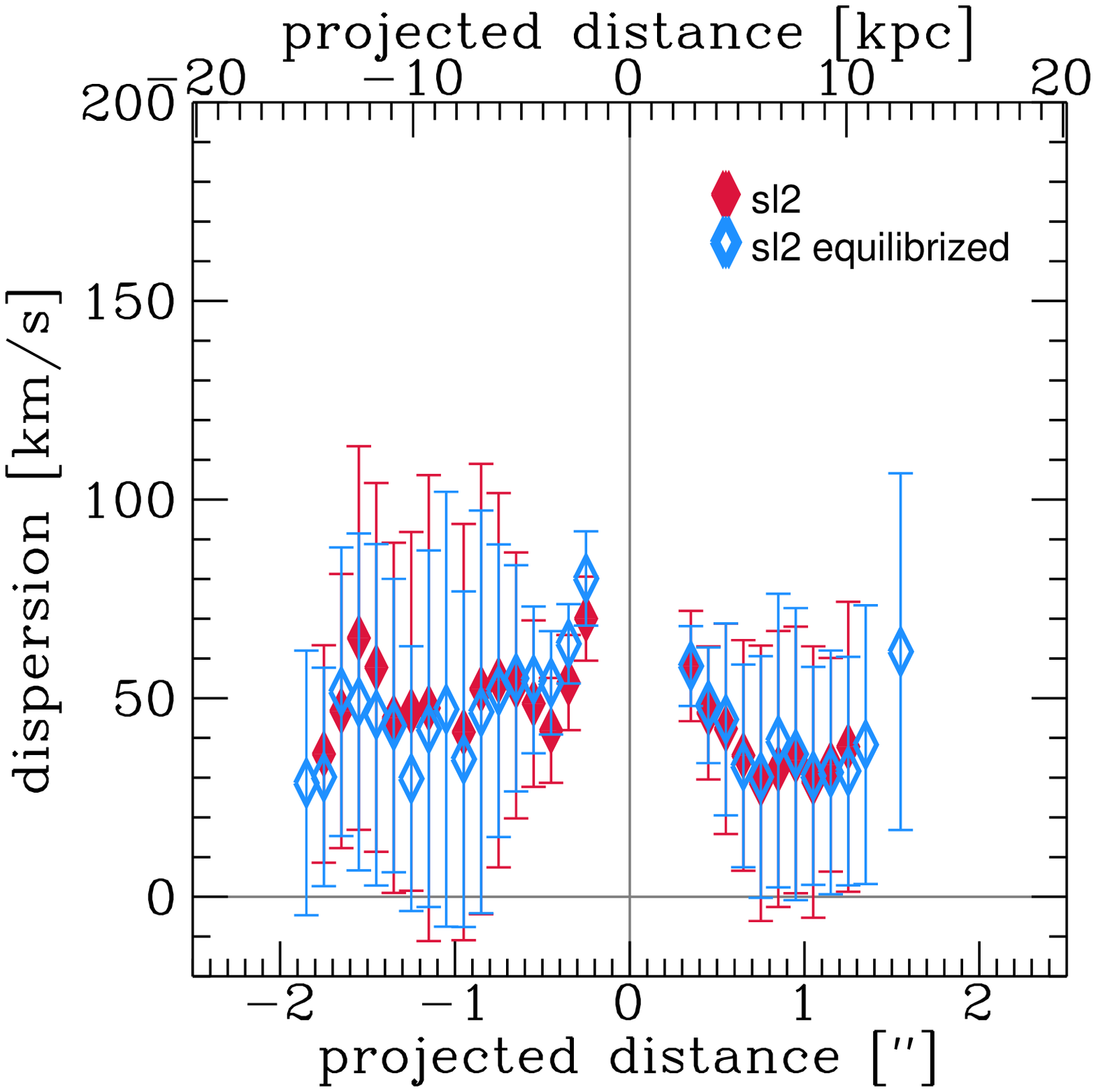}
	\caption[Effect of removing vertical motions II]{One-dimensional kinematic extractions from the mock data cubes before (red) and after (blue) removing vertical and radial velocity components for galaxy \#3, sightline 2. The rotation curve becomes more symmetric, and through increased $S/N$ extractions at larger galactocentric distances are possible. Effects on the velocity dispersion are minor.}
	\label{f:fiddling1d}
\end{figure}

We demonstrate the procedure and its effect by example of galaxy \#3, sightline 2, which (before equilibration) underestimates the central dark matter fraction by a factor of $\sim3$.
In Figure~\ref{f:fiddling2d} we show its projected two-dimensional kinematics before (left) and after (right) equilibration. The procedure particularly leads to more mirror-symmetric and smoother velocity maps.
In Figure~\ref{f:fiddling1d} we compare the corresponding one-dimensional kinematic extractions after creating mock data cubes for the processed galaxy with the original extractions. Note that for this exercise we use the identical noise cube in order to ensure a consistent comparison. In good agreement with the intrinsic two-dimensional kinematics shown in Figure~\ref{f:fiddling2d}, the fall-off on the receding side of the rotation curve is less extreme after equilibration, while the velocity dispersion along the kinematic major axis is not much affected. The more regular behaviour of the kinematics facilitates the line fitting and results in higher $S/N$ on average ($\Delta S/N\approx0.07$ for the full two-dimensional map, with $\Delta S/N\sim10$ in the center) such that the uncertainties on the extracted kinematics are slightly smaller, and extractions out to somewhat larger distances from the center are possible.

To quantify the gain in symmetry through equilibration, we compare the results of our asymmetry analysis (see Section~\ref{s:asym}). Using the overlapping coefficient, we find for the example shown in Figure~\ref{f:fiddling1d} an increase in symmetry from 0.41 to 0.54 (with 1 being completely symmetric, including uncertainties), i.e.\ by 32 per cent. If we consider our second method of fitting a quadratic function to one side of the rotation curve and calculating the reduced chi-squared statistics for the other side, we find a decrease from $\Delta\chi^2_{\rm red}=7.0$ to $\Delta\chi^2_{\rm red}=1.7$. Both methods show that the extracted rotation curve after equilibration is more symmetric.

Comparing the dynamical modeling results before and after equilibration for this example, we find that the baryonic parameters ($M_{\rm bar}$, $R_e$, $B/T$, $\sigma_0$) do not change beyond their $1\sigma$ uncertainties of the marginalized posterior distributions. However, the estimates of total dark matter mass and central dark matter fraction do: $f_{\rm DM}^v(<R_e)$ doubles from 0.09 to 0.18, and is therefore in better agreement with the intrinsic value, but still too low by a factor of $\sim2$. The estimate of the total halo mass increases from log$(M_{\rm halo}/M_{\odot})=11.7$ to $12.2$. This estimate is still lower than, but much closer to, the value of log$(M_{\rm halo}/M_{\odot})=12.7$ determined through the modified NFW fit to the intrinsic dark matter density distribution. This shows that the kinematic asymmetries caused by vertical and radial motions negatively affects the ability of our dynamical modeling to recover intrinsic values, particular with respect to dark matter.

More generally, the differences between the kinematic extractions along different lines of sight are reduced through the equilibration procedure. However, while more similar, there are still differences between different lines of sight that are larger than can be accounted for by uncertainties.
Overall, the effect of the equilibration technique on the regularity of the two- and one-dimensional kinematics underlines the impact of non-axisymmetric motions in the simulated galaxies.

\end{appendix}

\end{document}